\documentclass[10pt,journal,compsoc]{IEEEtran}

\ifCLASSOPTIONcompsoc
  \usepackage[nocompress]{cite}
\else
  \usepackage{cite}
\fi

\usepackage{graphicx}
\usepackage[normalem]{ulem}
\usepackage{color}
\usepackage{xcolor}

\usepackage{amsmath}

\usepackage{algorithm}
\usepackage{algpseudocode}

\usepackage{array}
\usepackage{booktabs}
\usepackage{colortbl}
\usepackage{multirow}

\ifCLASSOPTIONcompsoc
  \usepackage[caption=false,font=footnotesize,labelfont=sf,textfont=sf]{subfig}
\else
  \usepackage[caption=false,font=footnotesize]{subfig}
\fi

\usepackage{dblfloatfix}

\usepackage{url}

\begin{document}

\title{ReDy: A Novel ReRAM-centric Dynamic Quantization Approach for Energy-efficient CNN Inference}

\author{Mohammad Sabri,
        Marc Riera,
        and Antonio González
\thanks{Department of Computer Architecture, Universitat Polit\`ecnica de Catalunya (UPC), Barcelona, Spain, contact e-mails: mohammad.sabri@upc.edu, marc.riera.villanueva@upc.edu, antonio.gonzalez@upc.edu}%
}

\IEEEtitleabstractindextext{%
\begin{abstract}
Deep Neural Networks (DNNs) have achieved enormous success for a large variety of applications. The primary operation in DNNs is the dot product of quantized input activations and weights. Prior works have proposed the design of memory-centric architectures based on the Processing-In-Memory (PIM) paradigm. Resistive RAM (ReRAM) technology is especially appealing for PIM-based DNN accelerators due to its high density to store weights, low leakage energy, low read latency, and high performance capabilities to perform the DNN dot-products massively in parallel within the ReRAM crossbars. However, the main bottleneck of these architectures is the energy-hungry analog-to-digital conversions (ADCs) required to perform analog computations in-ReRAM, which penalizes the efficiency and performance benefits of PIM. To improve energy-efficiency of in-ReRAM analog dot-product computations we present ReDy, a hardware accelerator that implements a ReRAM-centric Dynamic quantization scheme to take advantage of the bit serial streaming and processing of activations. The energy consumption of ReRAM-based DNN accelerators is directly proportional to the numerical precision of the input activations of each DNN layer. In particular, ReDy exploits that activations of CONV layers from Convolutional Neural Networks (CNNs), a subset of DNNs, are commonly grouped according to the size of their filters and the size of the ReRAM crossbars. Then, ReDy quantizes on-the-fly each group of activations with a different numerical precision based on a novel heuristic that takes into account the statistical distribution of each group. Overall, ReDy greatly reduces the activity of the ReRAM crossbars and the number of A/D conversions compared to an static 8-bit uniform quantization. We evaluate ReDy on a popular set of modern CNNs. On average, ReDy provides 13\% energy savings over an ISAAC-like accelerator with negligible accuracy loss and area overhead.
\end{abstract}
}

\maketitle

\section{Introduction}\label{s:intro}
Modern machine learning applications such as image classification, speech recognition or machine translation are widely utilized on real-time systems including mobile devices and data centers. Deep Neural Networks (DNN) have demonstrated to be the most effective machine learning solution for a broad range of applications. DNNs are usually composed of hundreds of layers of artificial neurons where a single inference can demand billions of operations. The constant growth of DNNs makes them challenging to implement and run efficiently in conventional compute-centric architectures. The main performance and energy bottleneck of traditional von-Neumann architectures for these applications is the memory hierarchy due to the huge number of DNN inputs, weights, and partial outputs, resulting in numerous data movements which incur in much higher energy consumption than the operations~\cite{data_transfer_mobile, data_transfer_totall}.

In an attempt to enhance the performance and energy efficiency of DNN algorithms, several domain-specific architectures have been proposed over the last decade. However, due to the fast evolution and growing of DNN applications in terms of both computational complexity and memory requirements, current architectures suffer from various limitations. Even GPUs and popular DNN accelerators, such as the Google's TPU~\cite{TPU}, which have proven to be efficient hardware implementations for DNN inference and training, struggle to effectively provide the required memory bandwidth to run the latest DNN models. Consequently, a new architectural paradigm for data-intensive applications is required to fulfill the performance and memory bandwidth requirements of new machine learning algorithms and specially DNNs.

In order to address the problems of compute-centric architectures, and reduce the cost of communicating between memory and the processing units, a large number of works have proposed the design of a memory-centric architecture based on the so-called Processing-In-Memory (PIM) paradigm~\cite{MNEMOSENE, PUMA, FORMS, ISAAC, Direct-Current-Free, ResiRCA, XNOR-RRAM}. PIM consists in removing the need of transferring data to/from the processing units by executing the computations within the memory. The PIM approach is normally implemented by exploiting the analog characteristics of emerging Non-Volatile Memories (NVM) such as ReRAM crossbars~\cite {CASCADE, PRIME, ResiRCA, XNOR-RRAM}, Phase-Change Memory (PCM)~\cite{PCM}, or Spin-Transfer Torque Magnetic RAM (STT-MRAM)~\cite{STT}. Among all the promising memory technologies, ReRAM stands out in the implementation of DNN accelerators due to its high memory density to store the DNN weights, low read latency to access the data, and high performance PIM capabilities to perform the DNN dot-products~\cite{MPIM, Energy_Efficient, Boolean} within the ReRAM crossbars. In addition, since the leakage current of ReRAM is significantly lower than DRAM/SRAM cells, the energy efficiency can be highly improved.

ISAAC~\cite{ISAAC} and PRIME~\cite{PRIME} are popular PIM architectures for accelerating DNNs with ReRAM-based memory. At a high level, the core components of these accelerators are the ReRAM crossbars, the digital-to-analog converters (DACs) and the analog-to-digital converters (ADCs). ReRAM crossbars are used to store the DNN weights within the ReRAM cells, which in turn are used to perform the analog dot-product operations. DACs are employed to convert the input activations of each DNN layer to different voltage levels that are passed to the ReRAM crossbars to perform the computations. Finally, ADCs convert the analog outputs back to the digital domain. In order to reduce the cost of DACs, digital DNN activations are typically separated into units of 1-bit and streamed serially into the ReRAM crossbars to perform the analog dot-product operations. In this manner, only two voltage levels are needed to represent the inputs. In addition, activations of convolutional (CONV) layers from CNNs are commonly grouped according to the size of their filters and the size of the ReRAM crossbars. Then, different groups are processed in parallel. PIM DNN accelerators offer better performance and energy efficiency compared to GPUs and previous DNN accelerators. However, the area and power consumption of ADCs dominate the whole accelerator. For example, ISAAC’s ADCs account for 58\% of power and 31\% of area for a given tile of the accelerator~\cite{ISAAC}. Therefore, solutions to reduce and mitigate the costly A/D conversions are required to ease the execution of DNNs in ReRAM-based accelerators.

In this paper, we show how to efficiently exploit the bit-serial processing of groups of activations on CONV layers. We first propose a mechanism (ReDy) that dynamically quantizes groups of activations to different numerical precision. Compared to a fixed static quantization, a dynamic scheme allows to quantize at runtime with a finer-granularity, increasing the accuracy with lower bitwidths. Accordingly, the activity in the ReRAM crossbars, that is, the number of times the crossbars are employed to perform analog dot-products, as well as the number of A/D conversions, are highly reduced. The main challenge of this work is to determine on the fly which is the best precision for each group of activations with minimum hardware overhead and impact in accuracy. Our dynamic quantization scheme is based on the observation that each group of activations has a unique pattern and distribution of the values. We have experimentally observed that groups with a flatter distribution, that is, close to the uniform distribution, can be quantized to lower bitwidths with minor impact in accuracy. In addition, a subsample of the activations in each group is enough to approximate the distributions at runtime.

\begin{figure}[t!]
    \centering
    \includegraphics[width=1.00\columnwidth]{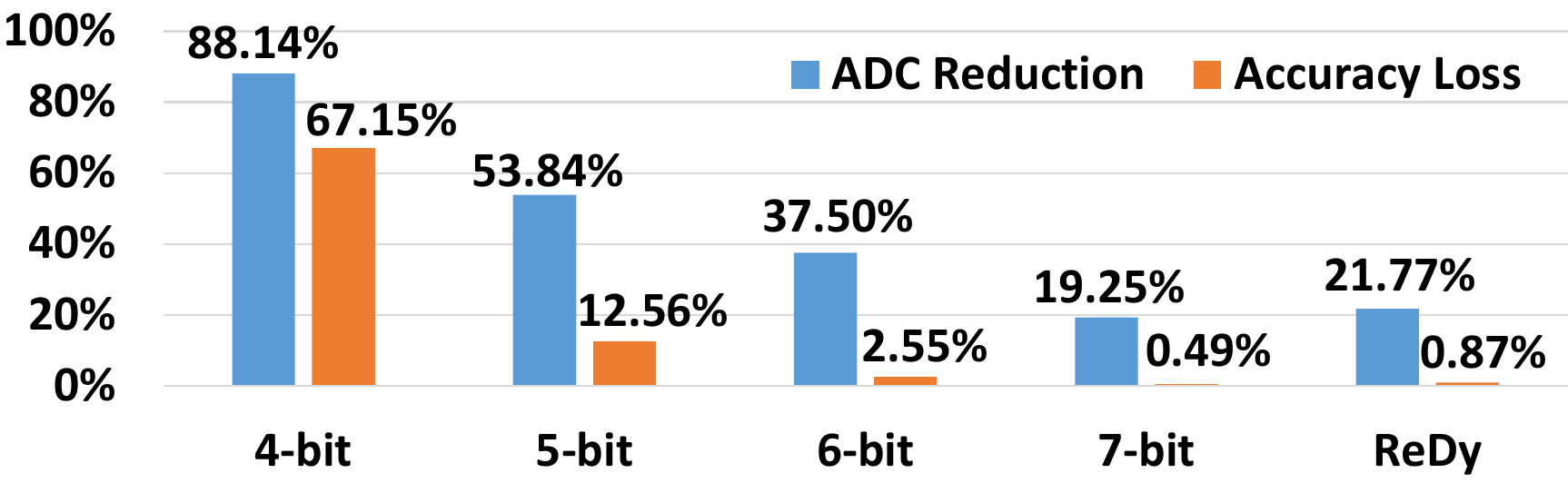}
    \caption{Reduction in A/D conversions and accuracy loss of ResNet-50 CNN for different static numerical precision uniform quantization and ReDy. Baseline is an ISAAC-like accelerator with 8-bit uniform quantization.}
    \label{fig:accuracyVSprecision}
\end{figure}

Figure~\ref{fig:accuracyVSprecision} shows the top-1 accuracy loss of ResNet-50~\cite{ResNet}, a popular CNN, when performing inference using different fixed bitwidth uniform quantization and our dynamic quantization scheme of the activations. In all cases, the CONV weights are statically quantized to 8-bits. Moreover, it also shows the reduction in A/D conversions for each numerical precision. As can be seen, our scheme (ReDy) is able to keep almost the same accuracy as the baseline that employs a fixed uniform quantization of 8-bits (i.e. under 1\% loss), while reducing the percentage of A/D conversions by 21.77\%, without retraining.

We also present a novel accelerator that implements the ReDy dynamic quantization scheme for CONV layers. ReDy is implemented on top of an ISAAC-like architecture, but it includes a new module to perform the dynamic quantization. The extra hardware required for our technique is modest since most of the components are already available in the baseline ISAAC-like architecture. ReDy only requires a small set of counters and comparators to compute the histogram of activations and some functional units to estimate the error. The extra elements represent a small increase in the area of the accelerator. Our experimental results show that the overheads are minimal compared to the savings in energy. ReDy not only avoids analog dot-product operations within the ReRAM crossbars, it also provides a significant reduction in analog-to-digital conversions. To summarize, this paper focuses on energy-efficient inference leveraging PIM ReRAM-based accelerators. The main contributions are:

\begin{itemize}

\item We analyze the distribution of different groups of activations and their effect on quantization error in multiple CONV layers. We observe that groups of activations with a flattened distribution can be quantized to lower bitwidths with minor impact in accuracy.

\item We propose a dynamic quantization scheme to exploit the bit-serial processing of groups of activations in ReRAM crossbars. Each group of activations is quantized to a different numerical precision based on the distribution of activations in the group and its similarity to a uniform distribution. On average, we quantize the activations of a popular set of CNNs to 6.1-bits incurring a negligible accuracy loss. Furthermore, this technique reduces the ReRAM crossbars activity and the number of A/D conversions by 33.8\%.

\item We present ReDy, a hardware accelerator that implements our dynamic quantization scheme. We evaluate ReDy for several CNNs by extending a PIM simulator named NeuroSim~\cite{Neurosim_trend}. ReDy reduces energy consumption by 13\% on average over an ISAAC-like accelerator.

\end{itemize}

The rest of the paper is organized as follows. Section~\ref{s:background} reviews the basics of CNN mapping and computation In-ReRAM. Section~\ref{s:discussion} discusses the observations on the distributions of groups of activations. Section~\ref{s:dq_method} describes the proposed ReDy algorithm. Section~\ref{s:hardware} details the hardware implementation of ReDy. Section~\ref{s:Methodology} presents the evaluation methodology and Section~\ref{s:Experimental_Results} discusses the experimental results. Finally, Section~\ref{s:Related_works} and Section~\ref{s:conclusion} sums up the related work and main conclusions.

\section{Background}\label{s:background}

\subsection{CNNs and CONV Layer}\label{subs:CNN_CONV}
Convolutional Neural Networks (CNNs) are a subset of DNNs commonly employed in the context of computer vision and image processing. CNNs are usually composed of four different types of layers: convolutional (CONV), fully-connected (FC), pooling, and normalization. CONV and FC are the most computationally and memory intensive layers, so are often targeted by accelerators and, particularly, ReRAM-based PIM architectures due to the high percentage of memory accesses and dot-product operations involved. The pooling layer consists of a down-sampling that applies a simple maximum or average function on a small window of activations. Finally, the normalization layer converts the activations to a set of values within a specific range and distribution.

A typical CNN in the computer vision domain includes several CONV layers to extract basic and complex features of the inputs, some pooling and normalization layers that are interleaved to decrease the size and mix multiple input feature maps into one and, finally, a few FC layers to classify the inputs to objects seen during training~\cite{ISAAC}.

\begin{figure}[t!]
    \centering
    \includegraphics[width=0.75\columnwidth]{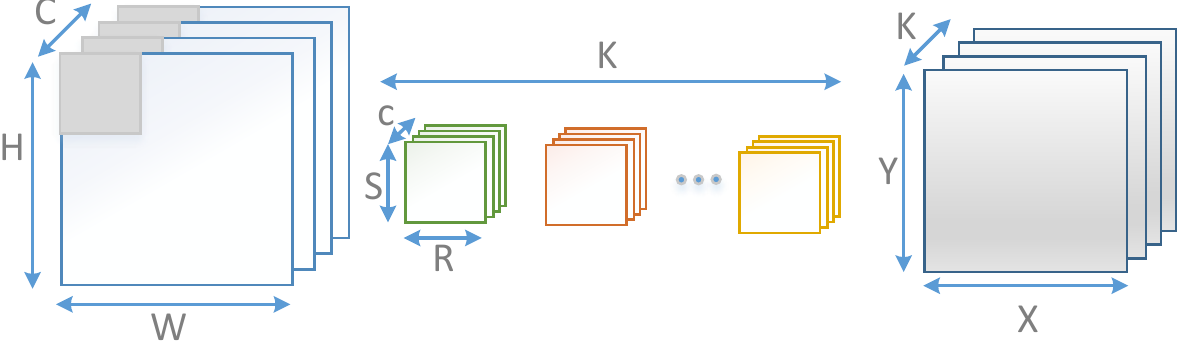}
    \caption{Example of a convolutional (CONV) layer.}
    \label{fig:convolution}
\end{figure}

An example of a CONV layer is shown in figure~\ref{fig:convolution}. An input feature map (ifmap) of size \(W \times H \times C\) is convolved with K kernels of size \(R \times S \times C\) to produce an output feature map (ofmap) of size \(X \times Y \times K\), where \(X = W - R + 1\) and \(Y = H - S + 1\). The kernels (i.e. weights) are learned through a training procedure~\cite{CASCADE}. A point \((x,y,k)\) of the ofmap, \(f^{out}(x,y,k)\), is calculated as follows:

\begin{equation}
\label{eqn:convolution}
f^{out}(x,y,k) =\sigma(\sum_{c=1}^{C}\sum_{r=1}^{R}\sum_{s=1}^{S} f^{in}(x+r,y+s,c) \times w_k(r,s,c)) 
\end{equation}

where \(f^{in}(x,y,c)\) is the input activation at position (x,y,c) of the ifmap, \(w_k(r,s,c)\) is the synaptic weight value of the \(k^{th}\) kernel at position (r,s,c), and \(\sigma\) is the activation function.

The FC layer can be viewed as a special case of a CONV layer, with many output feature maps, each using the largest possible kernel size, that is, the ifmap size is \(1 \times 1 \times C\), the ofmap size is \(1 \times 1 \times K\), and the kernels size is \(1 \times 1 \times C\).

\subsection{Uniform Quantization}
Quantization is a highly popular technique to map a continuous range of values to a discrete set. Equation~\ref{eqn:Quantization} shows an example of a function that quantizes real values (in floating-point, FP, precision) and maps them to an integer range.

\begin{equation}
\label{eqn:Quantization}
Q(r) = INT(r*s) - z 
\end{equation}

\begin{figure}[t!]
    \centering
    \includegraphics[width=0.75\columnwidth]{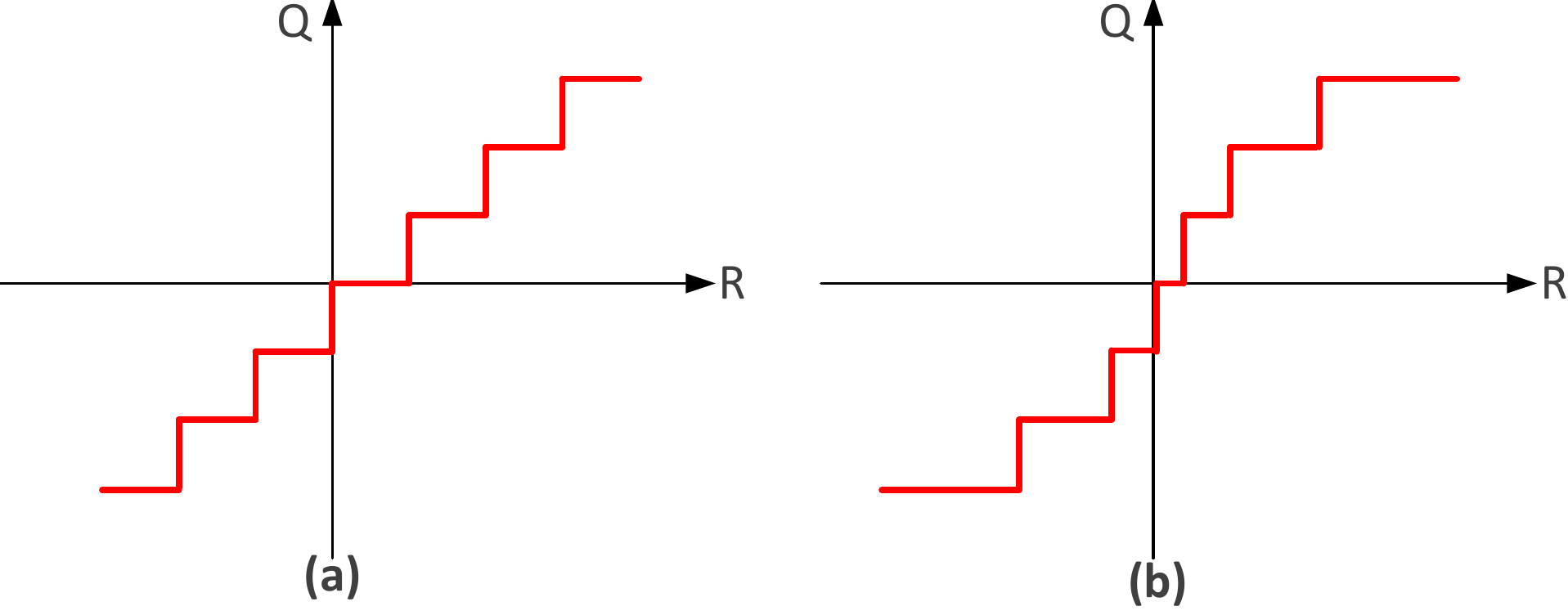}
    \caption{Comparison between (a) Uniform and (b) Non-Uniform quantization.}
    \label{fig:quantization}
\end{figure}

where \(Q(r)\) is the quantized value, \(r\) is a FP value, \(s\) is a scaling factor, and \(z\) is an integer offset. The \(INT\) function is a rounding to the nearest value. This method is also referred to as linear uniform quantization since the resulting quantized values (a.k.a. quantization levels) are uniformly spaced (Figure~\ref{fig:quantization}, left). Recently, non-uniform quantization methods have been proposed, but these require more complex computations that may not be as hardware friendly (Figure~\ref{fig:quantization}, right)~\cite{survey_Qauntization}. In this paper, we apply uniformly distributed linear quantization to the activations and weights of CONV and FC layers of CNNs. Then, we propose a dynamic quantization scheme for the activations of CONV layers.

Dequantization is often referred to as the procedure in which real values are recovered from the quantized values \(Q(r)\). Note that the recovered real values \~{r} will not be exactly equal to \(r\) due to the rounding operation. Consequently, the CNN may lose some accuracy. The accuracy loss tends to be negligible if enough quantization levels, which depends on the number of bits, are provided. Equation~\ref{eqn:Dequantization} describes the dequantization operation.

\begin{equation}
\label{eqn:Dequantization}
\tilde{r} = (Q(r) + z)/s
\end{equation}

\begin{figure}[t!]
    \centering
    \includegraphics[width=0.75\columnwidth]{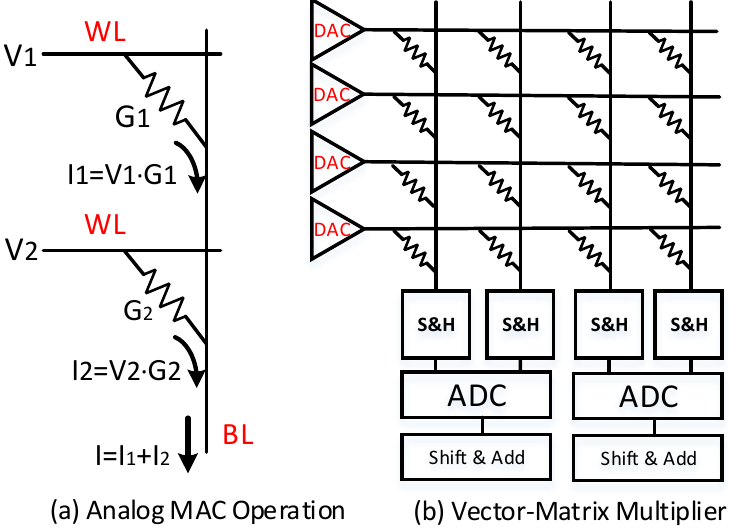}
    \caption{ReRAM Crossbar.}
    \label{fig:ReRAM_crossbar}
\end{figure}

\subsection{CNN Mapping and Computation In-ReRAM}\label{subs:CNN_Mapping}
Recent works~\cite{MNEMOSENE,CASCADE,ISAAC} have demonstrated the use of ReRAM devices to perform dot-product computations which, as described previously, are essential in many CNN layers. Figure~\ref{fig:ReRAM_crossbar} shows the structure of a ReRAM crossbar performing an analog MAC operation (a) and a Matrix Vector Multiplication (MVM) (b). In the ReRAM memory arrays, each bitline (BL) is connected to each wordline (WL) via a ReRAM cell that usually consists of a memristor with an access transistor, also known as the 1T1R cell design~\cite{marc}. Applying a voltage \(V_i\) to a ReRAM cell with conductance \(G_i\), results in a current \(V_i \times G_i\) passing from the cell to the bitline. Then, based on Kirchoff’s law, the total current in each BL is the sum of currents generated through all the cells that share the same BL. Thus, the total current (\(I\)) is the dot-product of input voltages at each row (\(V\)) and cell conductances (\(G\)) in a column as shown by Equation~\ref{eqn:MVM}. This can be used to encode the synaptic weights of each neuron as conductances of the ReRAM cells. Then, the total current of a BL is employed to compute the output of a neuron in a given layer. Note that the in-ReRAM analog dot-products can be done massively in parallel, allowing to perform a MVM in a single time step.

\begin{equation}
\label{eqn:MVM}
I =\sum_{i} V_i \times G_i 
\end{equation}

Figure~\ref{fig:ReRAM_crossbar}b also shows all the ReRAM crossbar peripherals involved in the analog dot-product operation for CNNs. Activations of each layer are converted to the analog domain through digital-to-analog converters (DACs). Converting N-bit inputs to proper distinguishable voltage levels requires costly DACs. Moreover, the converted analog voltages can be highly sensitive to external noise, affecting reliability and accuracy. To prevent these issues, previous works proposed to separate each input to units of 1 bit and stream them serially to the ReRAM crossbar. Thus, a simple DAC whose output only has two voltage levels is enough. Similarly, analog-to-digital converters (ADCs) are required to convert the analog dot-product results back to the digital domain. ADCs are costly in terms of area and power consumption, with an overhead that grows exponentially with the ADCs resolution. Consequently, current ReRAM-based architectures tend to include a small pool of ADCs shared among multiple crossbars. In addition, the ADCs bit resolution is selected so that the overheads are reduced with negligible impact in CNN accuracy. Other peripherals included are the Sample-and-Hold (S\&H) circuits which receive the BL current feeding it to a shared ADC unit when available. Lastly, due to current memristor technology limitations to represent multiple states safely, storing all the bits of a single weight in the same memory cell is not possible and, hence, multiple cells of the same row are used. Shift-and-Add units are employed to merge partial sums and calculate the final result of the dot-product operation.

\begin{figure}[t!]
    \centering
    \includegraphics[width=0.80\columnwidth]{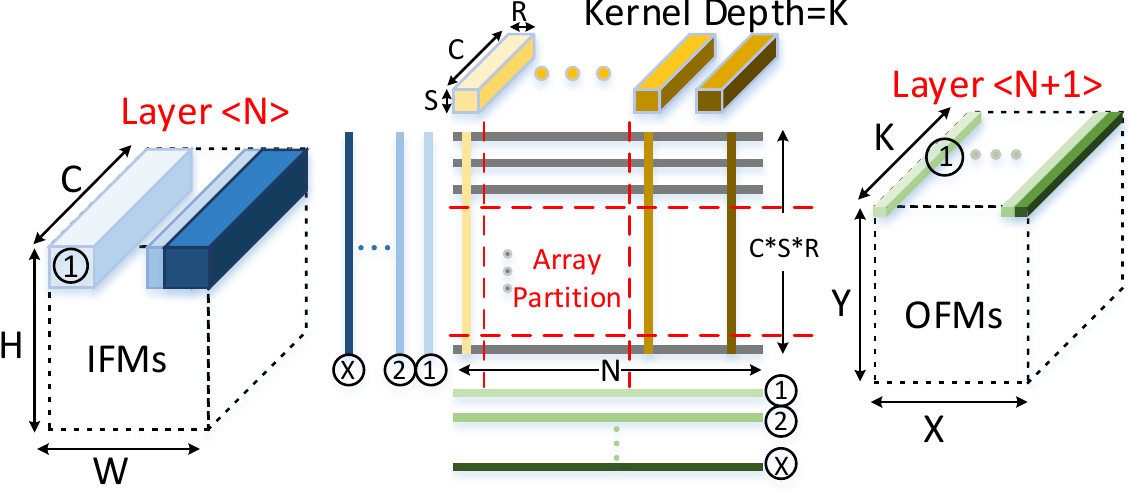}
    \caption{ReRAM Conventional CNN Mapping.}
    \label{fig:Conventional_Mapping}
\end{figure}

\begin{figure}[t!]
    \centering
    \includegraphics[width=1.00\columnwidth]{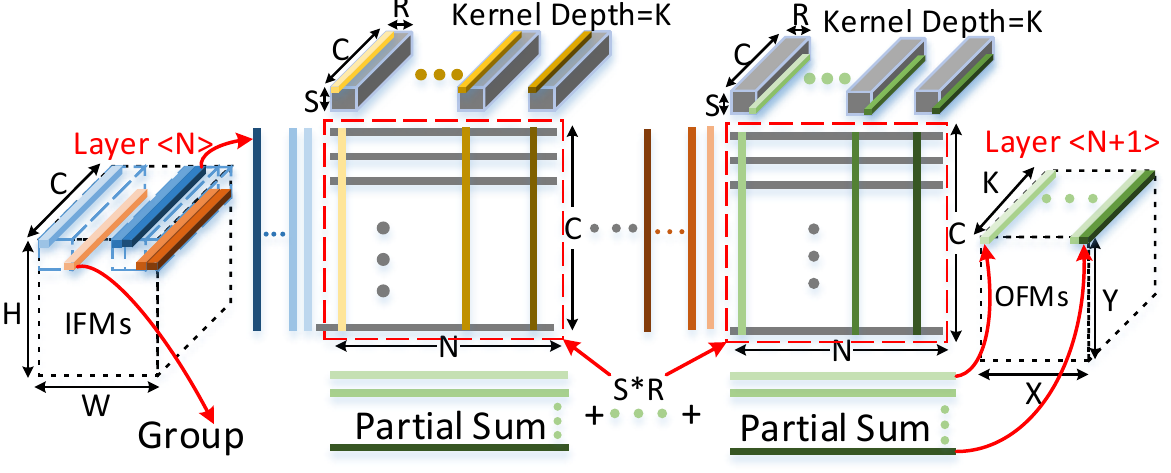}
    \caption{ReRAM Novel CNN Mapping.}
    \label{fig:Novel_Mapping}
\end{figure}

On the other hand, CNN mapping has a significant impact in the performance and energy consumption of ReRAM-based accelerators. Different dataflows have been proposed elsewhere~\cite{Neurosim_mapping,Training_map}. In the most conventional scheme, shown in Figure~\ref{fig:Conventional_Mapping}, each kernel of a CONV layer is unrolled into a single column of the ReRAM crossbar, which may be further partitioned into multiple arrays depending on the number of weights and the number of bits per weight. Then, a window of activations is also unrolled and applied to all the kernels in a single step. In the next step, the window slides over and the new activations are streamed bit-serially to the crossbar again. In each step of the convolutional operation, a percentage of the activations is reused from the previous iteration, wasting resources to access the same data. Considering the huge amount of dot-product operations in CONV layers, this frequent revisiting of activations from upper-level buffers can dramatically increase the latency and power consumption of the system. In order to maximize data reuse, the authors of NeuroSim~\cite{Neurosim_mapping} proposed a novel mapping scheme as shown in Figure~\ref{fig:Novel_Mapping}. Assuming that \(r,s,c\) are the coordinate axes of a kernel, weights along the c-axis (i.e. input channels) are mapped to the same column in the crossbar. Accordingly, each kernel is divided into different crossbars that produce several partial sums. By accumulating all of these partial sums, the final result of the dot-product is computed. In this work, we exploit NeuroSim's CNN mapping scheme to propose a novel dynamic quantization methodology, where a "Group" refers to activations that have the same (r,s) coordinates but different input channels.

\section{Analysis of CNN Activation Groups}\label{s:discussion}
In this section, we analyze the characteristics of CNN activation groups to introduce a criterion to perform efficient dynamic quantization. Determining the optimal numerical precision during the quantization process plays a vital role in reducing the induced error of the model. Accordingly, ReDy aims to determine the lowest possible numerical precision for activation groups to improve energy consumption and performance of ReRAM-based accelerators, while minimizing the quantization error to avoid a noticeable loss of accuracy. Generally, most previous works only take into account the range and its magnitude to perform uniform quantization, that is, a wider range may require more precision to represent all the values than a narrower range. However, this section shows that for deciding the optimal precision dynamically, the distribution of each group of activations is equally important to keep the critical information and reduce the amount of error.

\begin{figure}[t!]
    \centering
    \subfloat[]{\includegraphics[width=0.45\columnwidth]{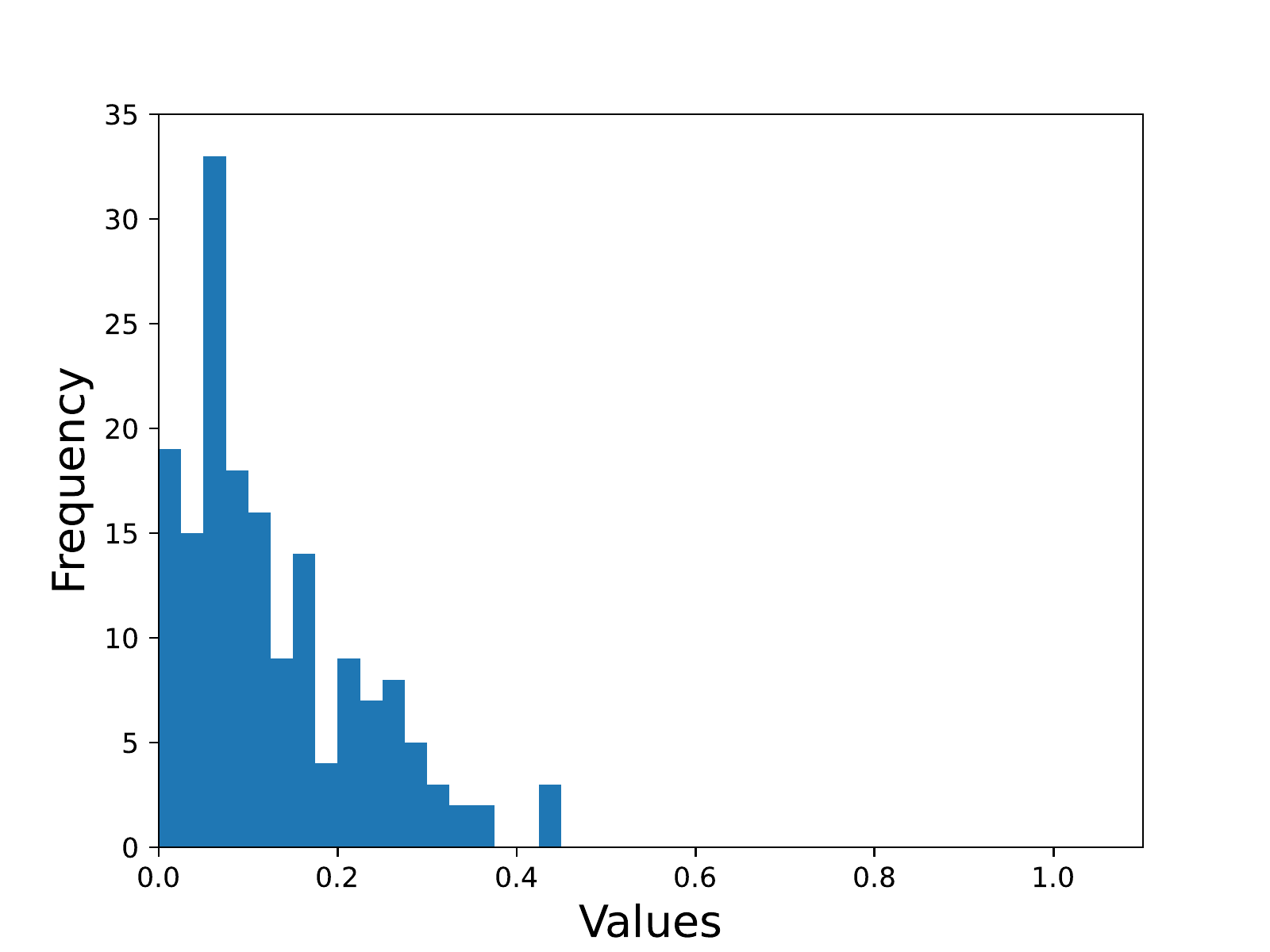}\label{fig:Congested}}
    \hfill
    \subfloat[]{\includegraphics[width=0.45\columnwidth]{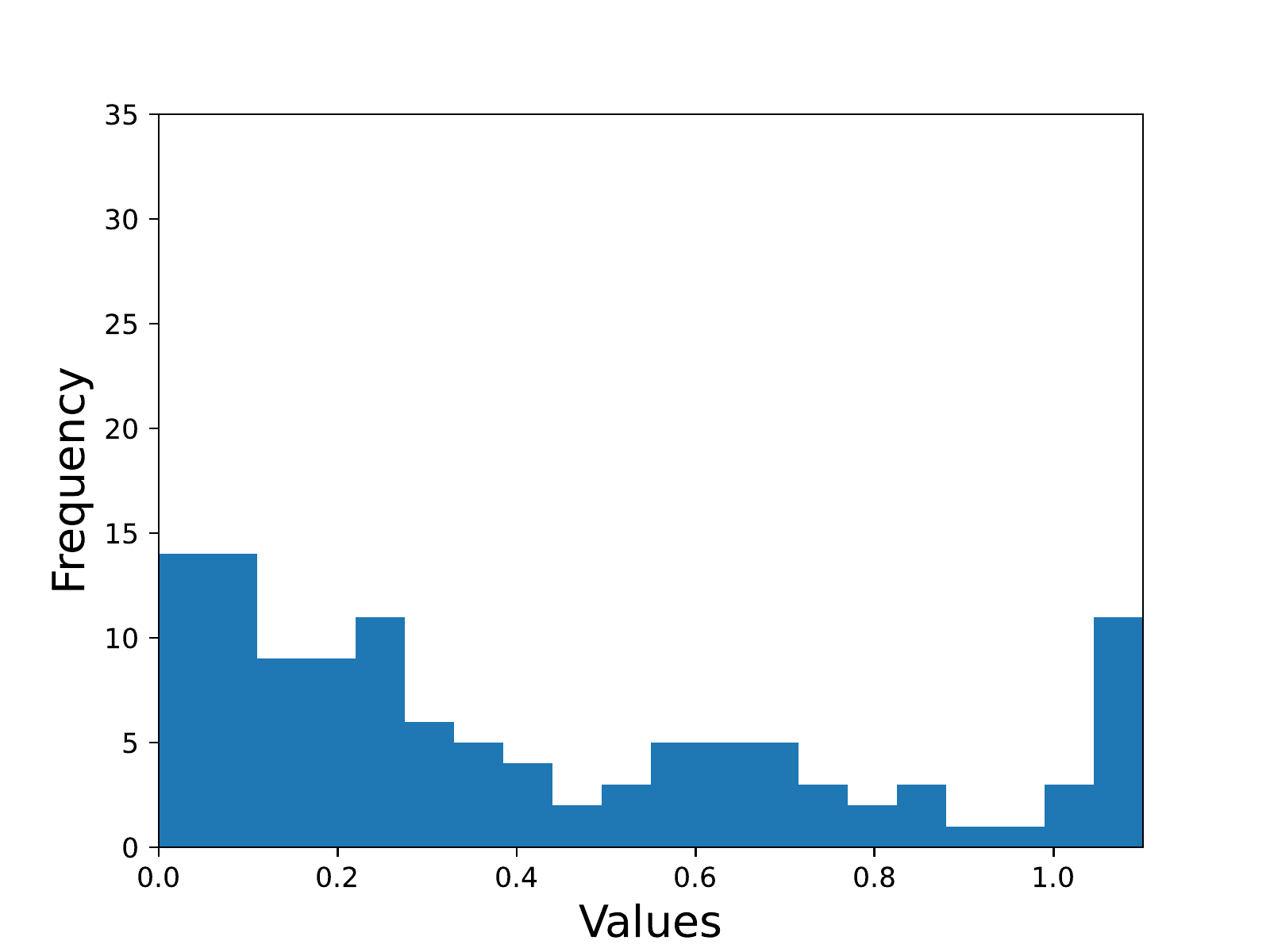}\label{fig:Less_Congested}}
    \caption{Distribution histogram of two different groups of activations for the same layer of the ResNet-50 CNN.}
    \label{fig:Distribution}
\end{figure}

As an example, Figure~\ref{fig:Distribution} shows the distribution histogram of two different activation groups from the same layer of the ResNet-50 CNN, where a group is defined as in Section~\ref{subs:CNN_Mapping}. As can be seen, the distribution varies significantly between different groups of activations. Figure~\ref{fig:Congested} presents the distribution of a group with a congested area close to zero. On the other hand, Figure~\ref{fig:Less_Congested} exhibits a flatter distribution with activation values spread more evenly within the range. In order to prove the importance of the distribution of CNN activation groups in the dynamic quantization process, we quantize uniformly these two groups to 4 and 8 bits. Then, we evaluate the induced error caused by the quantization scheme by measuring the Relative Root Mean Squared Error (RRMSE) metric.

The RRMSE is 1.67\% and 0.49\% for \ref{fig:Congested} and \ref{fig:Less_Congested}, respectively, when both groups are quantized uniformly to 8-bits. As expected, the group $(b)$ with the flatter histogram that is also closer to the uniform distribution yields a lower quantization error which can lead to less accuracy loss. Similarly, when decreasing the numerical precision of the uniform quantization to 4-bits, the RRMSE increases to 28.81\% and 7.89\% for groups $(a)$ and $(b)$ respectively, which corresponds to an increment of $17.2x$ and $16x$ respect the 8-bit uniform quantization. Based on these observations, we can potentially reduce the bitwidth of groups of activations with a more uniform distribution without affecting the accuracy of the network. Consequently, ReDy analyses the distribution of activation groups on-the-fly to determine the optimal numerical precision of the dynamic quantization process of each group.

\section{ReDy Quantization}\label{s:dq_method}
As previously described, state-of-the-art ReRAM-based accelerators for cognitive computing tend to include several crossbars to perform analog dot-products, as well as ADCs to convert the results from analog-to-digital domain. ReRAM crossbars and ADCs consume the vast majority of the total energy consumption in prior PIM accelerators~\cite{ISAAC}. Most of these accelerators employ 8/16-bit static uniform quantization to represent DNN weights and activations. As explained in Section~\ref{subs:CNN_Mapping}, DNN activations are divided into units of one bit and streamed serially to be processed in the ReRAM crossbars, reducing the DACs cost. The number of DNN activations and their numerical precision are key factors, proportionally dictating the activity within the ReRAM crossbars and ADCs.

For example, in a given CONV layer of DenseNet-161 with an input feature map of size $(56, 56, 192)$ and kernel size $(3, 3, 192, 48)$, the amount of crossbar activations and A-D conversions are $1.25E+09$ and $1.00E+10$ respectively, when all parameters are uniformly quantized with 8-bits, and the entire ImageNet validation dataset is executed. In contrast, a dynamic quantization scheme of the activations can reduce the ReRAM crossbars and ADCs activity to $9.03E+08$ and $7.22E+09$, that is, about 28\% reduction with minor effect on the network accuracy. Consequently, an efficient dynamic quantization approach where the numerical precision of the activations is determined with finer granularity (i.e. per activation) can provide substantial benefits to ReRAM-based accelerators for DNNs in terms of computational cost and energy consumption.

Leveraging a dynamic quantization scheme in traditional non-PIM DNN accelerators with conventional memory is complex since the functional units operate with a predefined fixed arithmetical precision and, hence, additional functional units supporting various bitwidths are required. Maximizing the utilization of all these functional units is a challenging task, and thus resources are likely to be wasted. Besides, in some cases the hardware required to quantize activations on-the-fly may be more costly than the benefits of computing with lower precision. On the other hand, the bit-serial streaming and processing of activations to perform analog dot-products is an intrinsic property of ReRAM-based accelerators, which allows a more natural implementation of a dynamic quantization scheme with varying numerical precision per activation. The main challenge is to determine the optimal numerical precision to achieve the best compromise between accuracy impact and energy consumption, which may require a small hardware overhead that should be easily compensated by the benefits of reducing activity in the crossbars and ADCs.

In this paper, we propose a ReRAM-centric Dynamic Quantization scheme of activations named ReDy to reduce the activity of ReRAM crossbars and ADCs. As discussed in Section~\ref{s:discussion}, the main potential features for dynamically deciding the optimal quantization bitwidth include the range, magnitude and distribution of the activations. In this work, we observe that the dispersion of the information distribution can be a highly beneficial criterion that provides a reasonable trade-off between numerical precision and quantization error. Accordingly, to select a suitable bitwidth for each activation, ReDy analyses the distribution of groups of activations where a group is defined as the activations along the input channel of a CONV layer as stated in Section~\ref{subs:CNN_Mapping}. Algorithm~\ref{alg:Dynamic_Quantization} shows the pseudo-code of the heuristic for determining at runtime the optimal numerical precision based on the statistical distribution of each group. ReDy is divided in three main steps marked in red.

\begin{algorithm}[t!]
\scriptsize 
\caption{ReDy Quantization Scheme}
\label{alg:Dynamic_Quantization}
\begin{algorithmic}[1]
    \Procedure{ReDy}{$a, b, id, UQ$} \Comment{in: $a$, $b$, $id$} \Comment{out: $UQ$}
    \If{$depth[id] < b$}
        \State $precision \gets 8$
    \Else
        \Statex {\hspace{2.75em} \textcolor{red}{\# Step 1}}
        \State {\(hist = Histogram(a, b, range[id])\)}
        \Statex {\hspace{2.75em} \textcolor{red}{\# Step 2}}
        \State {\(ED = \frac{depth[id]}{b}\)}
        \State {\(DU = \frac{1}{depth[id]}\sum_{i=0}^{b-1} \left| hist[i]- ED \right|\)}
        \Statex {\hspace{2.75em} \textcolor{red}{\# Step 3}}
        \If{\(p1 < DU \leq 2 \times \frac{b-1}{b}\)}
            \State $precision \gets 8$
        \ElsIf{\(p2 < DU \leq p1\)}
            \State $precision \gets 7$
        \ElsIf{\(p3 < DU \leq p2\)}
            \State $precision \gets 6$
        \ElsIf{\(p4 < DU \leq p3\)}
            \State $precision \gets 5$
        \ElsIf{\(p5  < DU \leq p4\)}
            \State $precision \gets 4$
        \Else
            \State $precision \gets 3$
        \EndIf
    \EndIf
    \State {\(UQ = UniformQuant(a, precision)\)}
    \EndProcedure
\end{algorithmic}
\end{algorithm}

Algorithm~\ref{alg:Dynamic_Quantization} is applied to each CONV layer $id$ of a DNN, where $a$ is a group of activations, and $b$ is the number of bins to generate the histogram of the distribution. The histogram is computed to determine the precision of all the activations in group $a$, all of which are serially streamed and processed simultaneously by a ReRAM crossbar as defined by the novel CNN mapping. The only exception is for layers where the number of input channels in the input feature map is small (i.e. less than $b$), such as the first CONV layer of most CNNs where the input channel size is just three (i.e. RGB). These layers may waste resources when the novel mapping is used, so the conventional mapping is employed instead for them. In addition, the histogram cannot be properly computed with small groups because there are not enough data to generate a statistically representative distribution. Consequently, ReDy performs a static 8-bit uniform quantization for these layers (lines 2,3), avoiding the quantization of the same input activations multiple times with different numerical precision.

In the first step of the algorithm, ReDy analyses the distribution of activations of a given layer by computing the histogram for each group (line 5). The histogram $hist$ is obtained by taking into account the pre-defined number of bins $b$ and the pre-computed range of the layer $range[id]$. For a given DNN, we obtain a global range of the activations per layer by considering the full training dataset in an offline procedure. Although groups can be quantized with different precision, the quantization range of the activations of a layer has to be the same because activations of different groups but same convolutional window are accumulated together. Therefore, after the dot-products related to a group are completed, the partial results are shifted to the maximum numerical precision supported so that the accumulation of results from different groups can be done in the same range. The histogram function accounts for the majority of the computations in the ReDy scheme. 

\begin{figure}[t!]
    \centering
    \includegraphics[width=0.85\columnwidth]{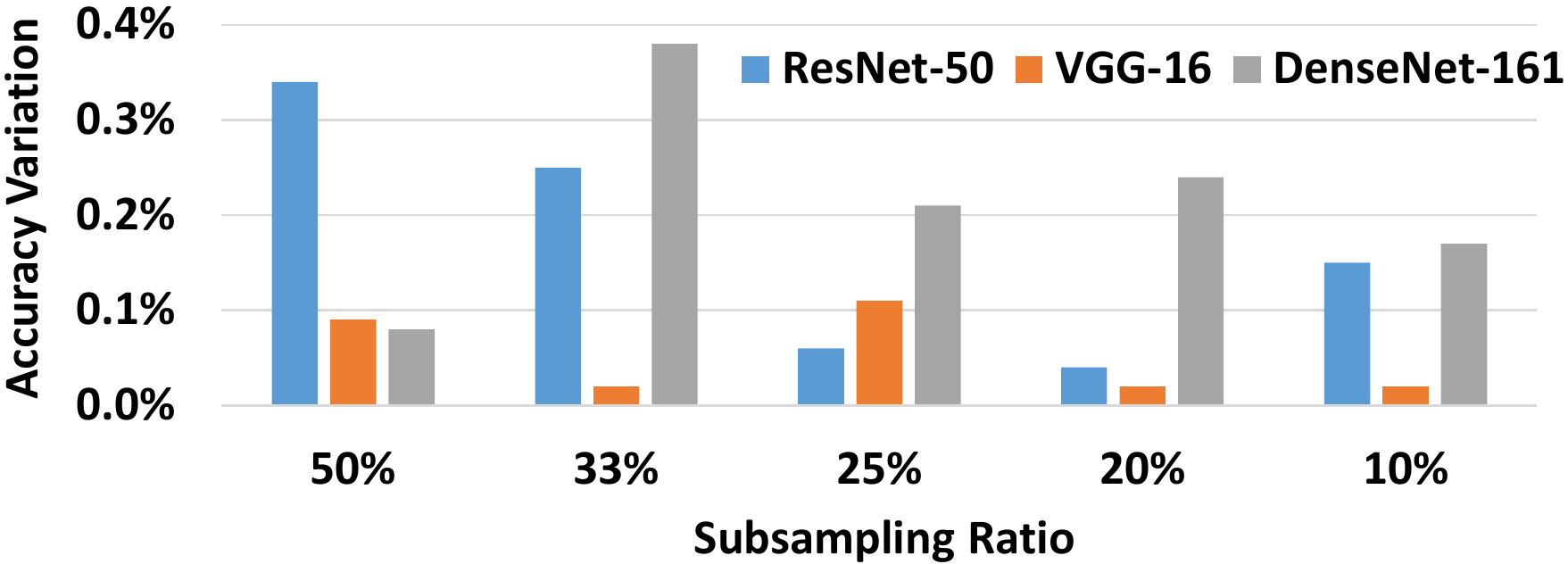}
    \caption{Accuracy variation across subsampling ratios.}
    \label{fig:subsampling}
\end{figure}

In order to reduce the cost of computing the histogram, ReDy leverages subsampling within a group to reduce the amount of activations that are compared and binned. However, subsampling may also impact the accuracy of the DNN since the computed histogram may not be representative of all the activations in the group, and the selected bitwidth may be suboptimal. Figure~\ref{fig:subsampling} shows the network model accuracy (Top-1) variation over different subsampling ratios for three different DNNs. Accuracy variation is defined as the absolute accuracy difference between ReDy without subsampling and ReDy with a given subsampling ratio. As can be seen, the accuracy does not change significantly across different subsampling ratios. In particular, 10\% subsampling ratio shows negligible accuracy variation while leading to a great reduction in the histogram computation cost.

\begin{figure}[t!]
    \centering
    \includegraphics[width=0.85\columnwidth]{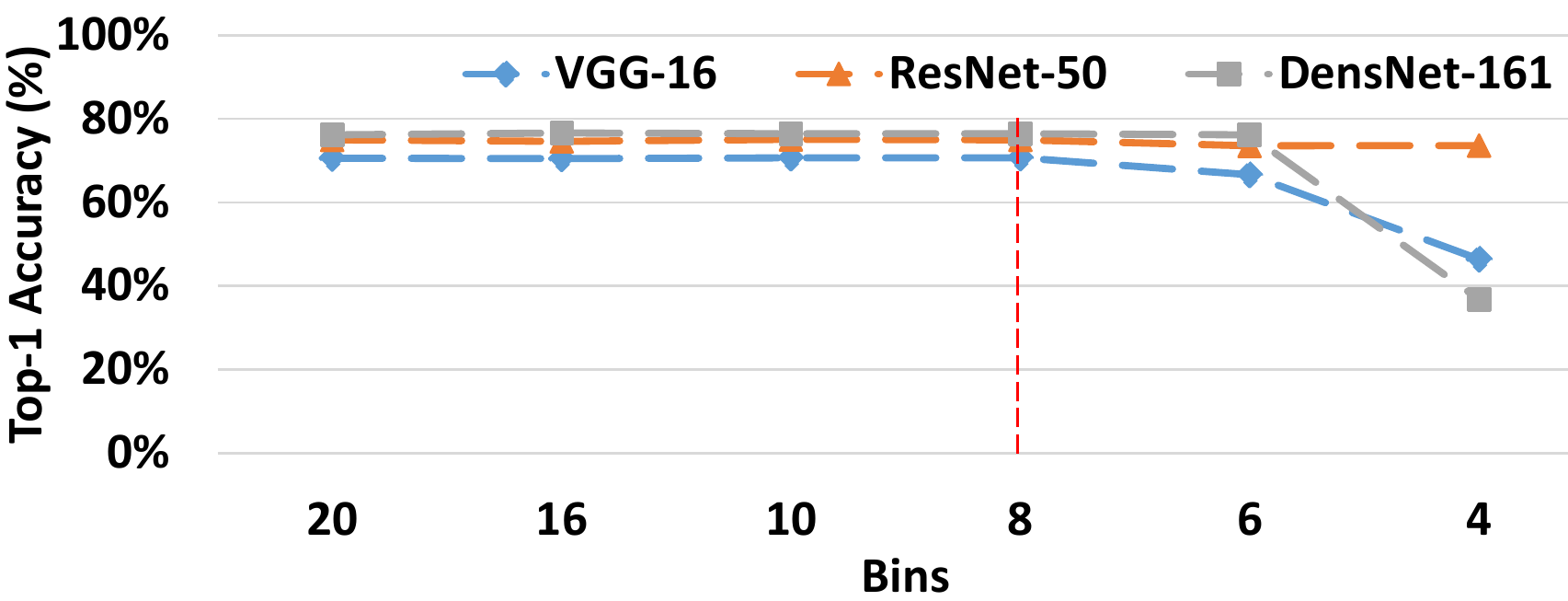}
    \caption{Sensitivity analysis on the number of bins.}
    \label{fig:optimal_bin}
\end{figure}

Similarly, regarding the number of bins $b$, we conducted a sensitivity analysis to assess the best compromise between accuracy loss and hardware overhead. From the hardware perspective, computing the histogram with less bins simplifies the hardware requirements by reducing the amount of comparators and counters. Therefore, the number of bins have an important impact on the latency and area overhead of ReDy. Figure~\ref{fig:optimal_bin} shows the effect of the varying number of bins on the accuracy of the networks. As can be seen, using 8 bins is the smallest configuration that has negligible impact on the accuracy and, thus, it is the one employed by ReDy to compute the histograms.

In the second step of the algorithm, ReDy evaluates how similar the histogram of the distribution of each group of activations is to the uniform distribution as a criterion for deciding the optimal bitwidth of the group (lines 6-7). As observed in Section~\ref{s:discussion}, a congested area in the distribution, that is, a bin with a frequency much bigger than the rest may require higher numerical precision to be represented with low quantization error. On the other hand, a flatter and smoother distribution may allow a reduction in precision with low impact on error. The expected distribution (\(ED\)) denotes the number of expected samples in each bin when the group distribution is perfectly uniform, that is, \(ED\) is calculated by dividing the group size by the number of bins. Note that the group size is the depth of the input feature map (i.e. activation channels), which may vary from layer to layer. The deviation from uniform (\(DU\)) metric quantifies the amount of similarity to the uniform distribution, which is computed by measuring the mean absolute error (MAE) of \(hist\) and \(ED\). \(DU\) is zero if the actual distribution exactly matches the uniform distribution (flattest distribution), while in the worst case \(DU\) will be \( 2 \times \frac{b-1}{b}\) for a highly congested distribution which has all the points in one bin.

Finally, in the last step, the numerical precision of each group is determined by comparing \(DU\) against a set of \(p_i\) coefficients (lines 8-20). The bitwidth can vary from 3-bits to 8-bits based on the similarity to the uniform distribution. The \(p_i\) coefficients are obtained empirically depending on the quantization error tolerance of each network. Finally, each group of activations is quantized uniformly based on the selected precision (line 22).

\section{ReDy Accelerator}\label{s:hardware}
This section describes the hardware support required to implement ReDy. First, we present the main hardware components of the ReDy accelerator for CNN inference. Next, we show how CONV layers are executed in the accelerator using ReDy. Finally, we describe how to efficiently scale the design and support other types of layers and networks.

\subsection{Architecture}\label{subs:General ReRAM-Based Architecture}
\begin{figure}[t!]
    \centering
    \includegraphics[width=0.75\columnwidth]{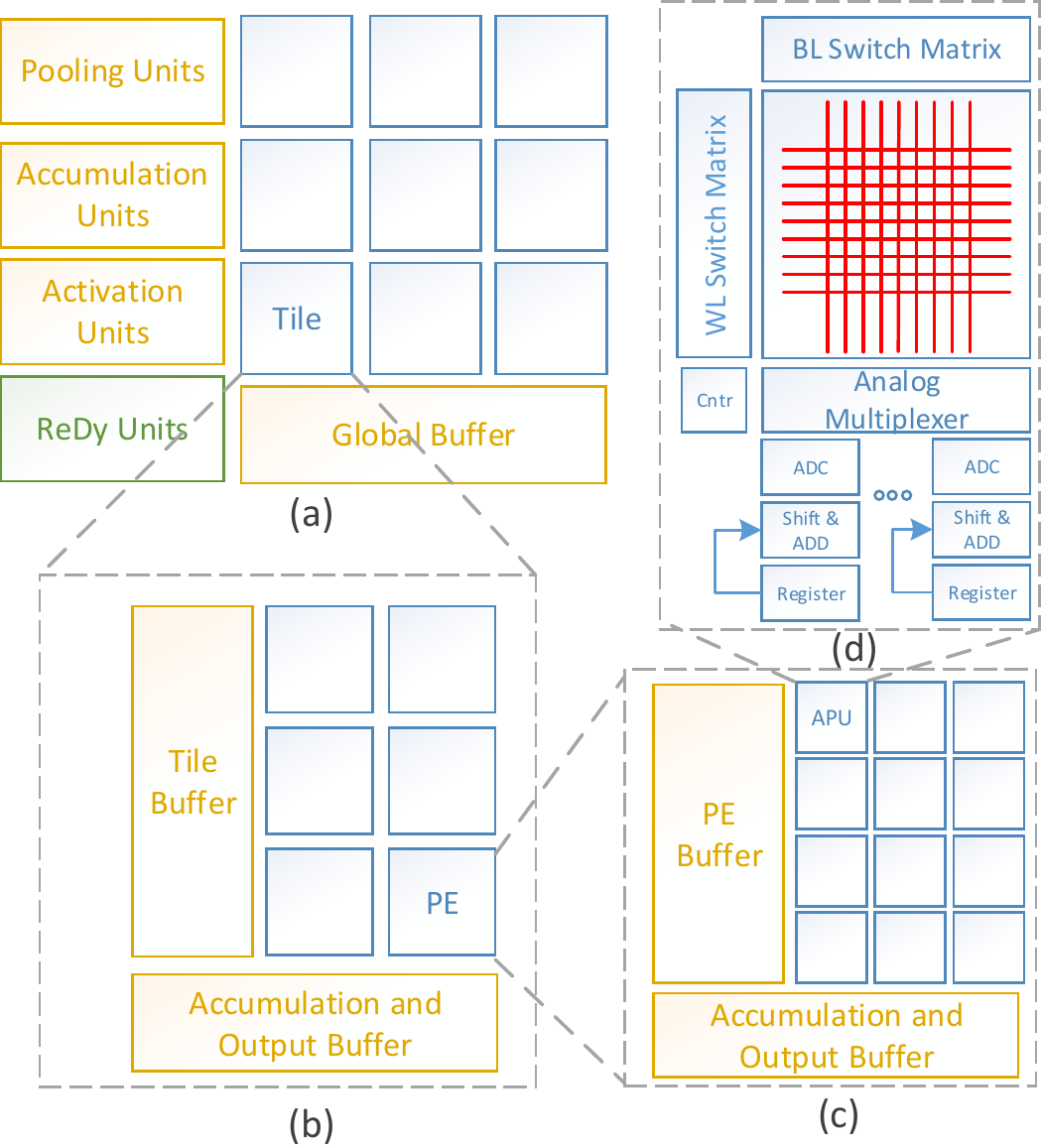}
    \caption{Architecture of the ReDy accelerator including the organization of a single (a) Chip, (b) Tile, (c) Processing Element (PE), and (d) Analog Processing Unit (APU). Components colored in green represent the required extra hardware.}
    \label{fig:hierarchy}
\end{figure}

In this section, we present an accelerator that takes advantage of the ReDy scheme to dynamically quantize activations in the CONV layers of different CNNs. ReDy exploits the bit-serial processing of activations to save analog dot-products and A/D conversions. Figure~\ref{fig:hierarchy} illustrates a high-level schematic of the top-down architecture of the accelerator. The main blocks of the hierarchy include a Chip, Tiles, Processing Elements (PE), and Analog Processing Units (APU). This tile-based architecture is inspired in ISAAC~\cite{ISAAC}, which has proven to be an efficient PIM accelerator to perform DNN inference.

Figure~\ref{fig:hierarchy}a depicts a high-level view of the chip architecture. A single chip is composed of multiple tiles, a global buffer to store activations, pooling and activation (sigmoid or ReLU) units, our ReDy units to determine the numerical precision of activation groups, and accumulation units to add all partial results from different tiles. Likewise, Figure~\ref{fig:hierarchy}b presents the structure of a tile, which contains several PEs, a tile buffer to store input activations, an output buffer to store partial sums, and accumulation modules to add the partial sums from different PEs. Similarly, as Figure~\ref{fig:hierarchy}c shows, a PE is built up by a group of APUs, a PE buffer, an accumulation unit, and an output buffer.

Finally, Figure~\ref{fig:hierarchy}(d) shows the main components of an APU consisting of a crossbar array to store synaptic weights that is built based on ReRAM cells (1T1R). An APU also includes a BL switch matrix to initialize the bitlines, a WL switch matrix to stream input activations to the ReRAM crossbar, an analog multiplexer, and a shared pool of ADCs, shift registers and adders to accumulate and shift the partial results from different BLs and iterations. More details on the analog dot-product operation with ReRAM crossbars can be found in Section~\ref{subs:CNN_Mapping}.

\begin{figure}[t!]
    \centering
    \includegraphics[width=1.00\columnwidth]{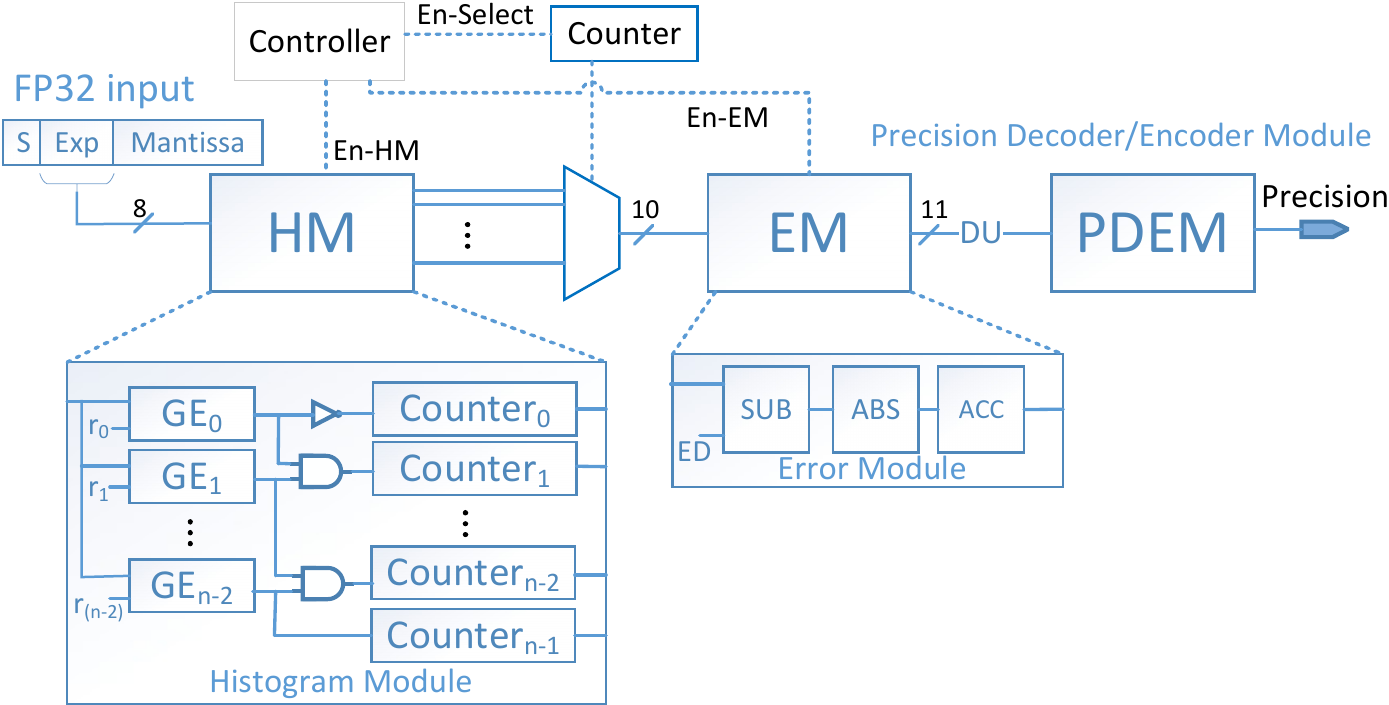}
    \caption{ReDy Quantization Unit Architecture.}
    \label{fig:Proposed_hardware}
\end{figure}

Figure~\ref{fig:Proposed_hardware} presents the hardware architecture of the proposed scheme for determining the quantization bitwidth of each group of activations. Similar to Algorithm~\ref{alg:Dynamic_Quantization}, the ReDy Unit is split into three different modules; the Histogram Module (HM), the Error Module (EM), and the Precision Decoder/Encoder Module (PDEM). In addition, the controller of the unit orchestrates all the operations by enabling the corresponding modules.

The first module, HM, is in charge of calculating the histogram for representing the distribution of the group of FP32 activations. However, instead of comparing FP32 values, which is costly from the hardware perspective, the HM extracts the integer exponents of the FP activations, and compares them with the boundaries $r$ of the histogram bins. These boundaries are obtained offline by dividing the global range of a given layer by the number of bins. However, the boundaries are also weighted to account for the non-linear effect of the exponents. The exponents allow to understand and classify the magnitude of the FP values without losing any accuracy. Then, for each bin, there is a counter that is increased whenever the resulting comparison evaluates to true, that is, each activation in the group is sampled into the corresponding bin of the histogram. The HM hardware is composed of a set of $N-1$ comparators, $N$ counters, and a few logic gates, where $N$ is the number of bins in the histogram.

Once the histogram of a group of activations is obtained, the Error Module (EM) starts computing the Deviation from Uniform (DU) metric, which is defined as the divergence of the histogram with respect to the Expected Distribution (ED). As described in Section~\ref{s:dq_method}, the ED of each layer is computed offline and loaded by the controller when needed. Similar to the boundaries, the ED values are also non-linear. The controller iterates over the HM counters storing the number of samples in each bin of the histogram using a multiplexer and a small counter. In each iteration, the EM module accumulates the absolute error between the histogram bin value and ED as shown in Algorithm~\ref{alg:Dynamic_Quantization}. Finally, the mean absolute error (MAE) is computed by an implicit division of the accumulated value by the group size (depth), that is, a fraction point is assumed to be placed in the $log_{2}(depth)$-th bit, resulting in the DU metric, which is sent to the third and last module. The EM hardware is composed of a set of functional units to perform subtraction (SUB), absolute (ABS), and accumulation (ACC) operations.

Finally, the Precision Decoder/Encoder Module (PDEM) is responsible for determining the precision of a given group of activations. The numerical precision is decided by considering the DU value computed in the previous module and the $p1$ to $p5$ coefficients. As described in Section~\ref{s:dq_method}, the coefficients represent reasonable error thresholds that take into account the best trade-off between numerical precision and CNN accuracy. These coefficients are computed offline and loaded into the accelerator during its initialization. The PDEM module is mainly composed of a set of five comparators to contrast the DU with the coefficients and determine the most suitable precision according to the results of the comparisons.

\subsection{Dataflow}\label{subs:execution_flow}
\begin{figure}[t!]
    \centering
    \includegraphics[width=0.50\columnwidth]{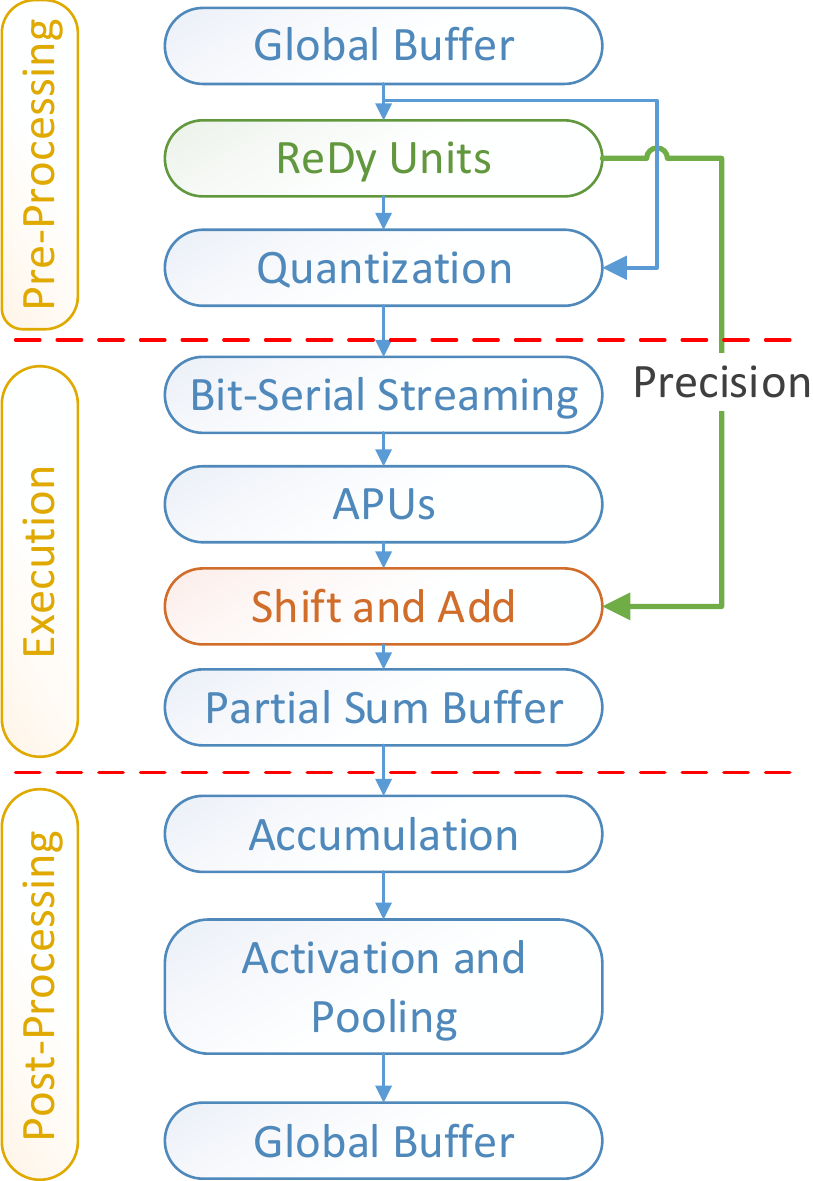}
    \caption{ReDy Execution Dataflow. Steps colored in green/red are added/modified to the execution flow.}
    \label{fig:execution_flow}
\end{figure}

We enhance the dataflow of the ISAAC-like accelerator to integrate our ReDy scheme in a ReRAM-based architecture. Figure~\ref{fig:execution_flow} illustrates the modified dataflow of the ReDy accelerator with a flowchart. The additional steps are marked in green color similar to the new ReDy component added to the baseline architecture. Besides, steps of the flowchart that are not new but have been changed are marked in red color. The proposed dataflow includes three main stages: Pre-Processing, Execution, and Post-Processing.

In the \emph{Pre-Processing} stage, groups of activations are fetched from the global buffer and distributed into tiles based on each layer floor-planning. Similar to previous works~\cite{ISAAC, dynamic_reram, Neurosim_mapping}, ReDy considers that the on-chip memory of the accelerator is sufficient to store the synaptic weights of the entire CNN, which are mapped as described in Section~\ref{subs:CNN_Mapping}. Depending on the size of each layer and the hardware configuration of the accelerator, a set of tiles is assigned to a given layer and, hence, groups of activations are divided accordingly across different tiles. Later on, the partial dot-products of different tiles will be accumulated to obtain the final results. Before distributing the activations, these are quantized with our ReDy scheme. For a given layer, activations are read from the global buffer, grouped, and sent to the ReDy units to determine the individual numerical precision of each group. Then, groups of activations are uniformly quantized by their corresponding precision. In contrast, the baseline dataflow simply quantizes to 8/16-bits statically. In subsequent steps, the quantized groups are distributed to the lower levels of the hierarchy until reaching the APUs that store the corresponding weights of the layer in their ReRAM crossbars. Note that the precision of each group is sent along the activations.

In the \emph{Execution} stage, each group is first converted into a bit stream and applied to the crossbars of the APUs to perform the analog dot-products. Next, ADCs capture the analog partial results and convert them into digital values. Note that each weight can be stored using multiple consecutive ReRAM cells due to technology limitations. Therefore, the values of several BLs are shifted and added to produce the partial sum of an output feature map. Similarly, a group of activations is processed serially so the results of each iteration are shifted and added as well. Then, partial sums of different activation groups but same convolutional window are scaled to the same range, that is, the range corresponding to the maximum precision supported, to accumulate them together. No additional hardware is required for this operation since the baseline APUs already have the components to scale the partial results.

Finally, in the \emph{Post-Processing} stage, the partial results obtained in different APUs are forwarded and added together when needed in the accumulation units of the upper levels of the hierarchy. As explained above, the main reason is that a given layer may be distributed among multiple APUs, PEs, and Tiles. Therefore, the partial sums of the deeper levels are propagated to the higher levels to continue the accumulations and produce the final output feature map results. At the chip level, the final results are dequantized, and the activation and pooling units are triggered to perform the corresponding operations before storing the values back to the global buffer.

In summary, all layers of a DNN are executed concurrently in a deep pipeline similar to ISAAC~\cite{ISAAC}. Each group of activations is pre-processed by uniformly quantizing them to a different numerical precision determined by ReDy. Then, groups are executed bit-serially within ReRAM crossbars of the APUs. Finally, results are post-processed and re-distributed again to the corresponding tiles to continue the pipelined execution. ReDy does not incur any latency overhead compared to the baseline dataflow since the longest stage of the pipeline is still the execution, which can be overlapped with the other stages.

\subsection{Design Flexibility}\label{subs:Design_Flexibility}
\begin{figure}[t!]
    \centering
    \includegraphics[width=0.50\columnwidth]{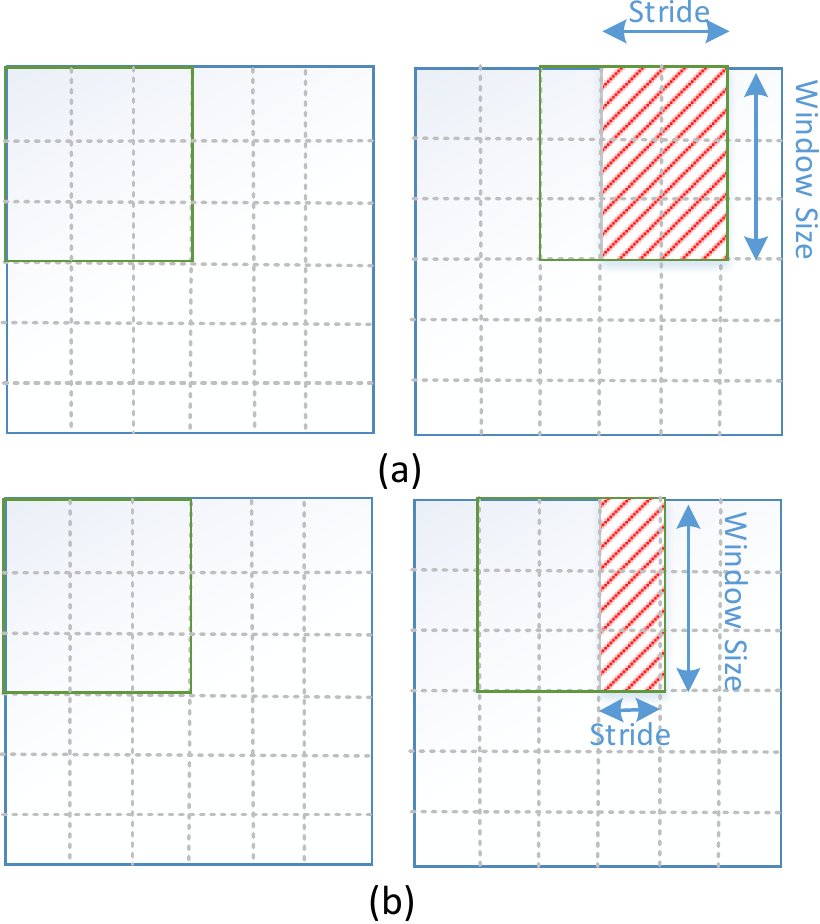}
    \caption{Example of the scalability of ReDy for a CONV layer when a) stride=2 and b) stride=1.}
    \label{fig:stride}
\end{figure}

A key parameter in the dataflow of ReDy and its hardware implementation is the number of ReDy units. These are required to determine the quantization bitwidth of multiple groups of activations in parallel since all the layers of the network are executed simultaneously. The number of ReDy units affects the flexibility and scalability of the accelerator. If there are not enough units the latency of our scheme cannot be completely hidden by other stages of the pipeline reducing the performance. On the other hand, more units increase the area of the accelerator. We take into account the stride of each CONV layer of a given CNN to decide on the optimal number of ReDy units. Figure~\ref{fig:stride} shows an example of the number of ReDy units required for a CONV layer with different strides. In each iteration of a convolution, the window slides over different groups of activations whose precision should be determined before quantizing and dispatching to the tiles. As can be seen, at least \(Stride \times WindowSize\) ReDy units are needed to avoid stalls during the execution of the layer. In most CNNs, the stride of the CONV layers tends to be small (e.g. 1-4) and, hence, the number of ReDy units required and their overhead is minimal compared to the rest of the hardware components of the accelerator. Consequently, by adjusting the number of units, ReDy can keep the performance of the accelerator for any CNN with negligible area overhead. More details on these overheads are provided in a later section.

Although ReDy is specialized in efficient inference of CONV layers, it can also support fully-connected (FC) layers with minor benefits from our dynamic quantization scheme. As described in Section~\ref{subs:CNN_CONV}, a FC layer can be viewed as a special case of a CONV layer where the ifmap size is \(1 \times 1 \times C\). Thus, for each execution of a FC layer there is only one group of activations that can be quantized to the numerical precision determined by ReDy. This makes our scheme less efficient because the probability of quantizing a single group with many activations at low precision is smaller. For example, VGG-16 quantizes 76\% of the FC activations to 6-bits while all of ResNet-50 FC activations are quantized to 7-bits. We plan to optimize the inference of FC layers with ReDy in future work with an scheme that divides the activations in smaller groups.

\section{Methodology}\label{s:Methodology}
We have extended NeuroSim~\cite{Neurosim_github}, a trace-based simulator for cognitive computing PIM accelerators, to accurately model two different systems: an ISAAC-like accelerator and our ReDy scheme as presented in Section~\ref{s:hardware}. NeuroSim models a PIM ReRAM-based architecture and provides a summary of the accelerator main features including area, throughput and power. Table~\ref{tab:Param} shows the common parameters for the baseline accelerator and our scheme. The baseline configuration is an ISAAC-like architecture which employs ReRAM crossbars to perform analog dot-products. In addition, inputs and weights are quantized to 8-bit integers without any accuracy loss for our set of CNNs. The mapping procedure of CNNs to ReRAM crossbars is vital to guarantee high-enough memory utilization. Memory utilization is defined as the portion of ReRAM cells that are programmed with synaptic weights, since a suboptimal configuration may lead to wasted ReRAM cells. Consequently, NeuroSim developed an algorithm to automatically generate the best floorplan of the accelerator to optimize memory utilization. In particular, NeuroSim determines the amount of APUs in a PE, the quantity of PEs in a Tile, and the number of Tiles in a Chip, based on multiple user-defined parameters such as the crossbar array size and the CNN model structure. In addition, NeuroSim selects the optimal clock frequency of the accelerator taking into account the estimated maximum latency of the analog dot-products within a ReRAM crossbar. Similar to ISAAC, we set the crossbar size in each APU to $128 \times 128$ ReRAM cells, where each cell has a 2-bit resolution. Thus, 8-bit weights are represented with 4 consecutive cells.

\begin{table}[t!]
\caption{Parameters for the baseline and ReDy accelerators.}
\label{tab:Param}
\centering
\resizebox{0.70\columnwidth}{!}{%
    \centering
    \begin{tabular}{|c|c|}
    \hline
    \cellcolor[gray]{0.9} Technology & 32 nm \\
    \cellcolor[gray]{0.9} Frequency & 400 MHz \\
    \cellcolor[gray]{0.9} Chip/Tile/PE Buffers Type & SRAM \\
    \cellcolor[gray]{0.9} Crossbar Memory Cell Type & ReRAM \\
    \cellcolor[gray]{0.9} Crossbar Array Size & 128*128 \\
    \cellcolor[gray]{0.9} Crossbar Memory Cell Precision  & 2-bits \\
    \cellcolor[gray]{0.9} Weights Precision & 8-bits \\ 
    \cellcolor[gray]{0.9} Number of ADCs per Crossbar & 16 \\
    \cellcolor[gray]{0.9} ADCs Sampling Precision & 5-bits \\
    \hline
    \end{tabular}%
}
\end{table}

On top of the baseline architecture, we have implemented our novel scheme by adding the new ReDy units. As described in Section~\ref{subs:Design_Flexibility}, the number of units is determined based on the stride parameter of each CONV layer of a given CNN. Regarding area, latency and energy consumption of the ReDy units, the additional hardware components are implemented in Verilog, and synthesized to obtain the delay and power using Design Compiler~\cite{Design_compiler}, the modules of the DesignWare library, and the technology library of 28/32nm from Synopsys. For the technology library, we use the standard low power configuration with 0.78V. Moreover, ModelSim~\cite{ModelSim} is used to verify the design. The results obtained with the aforementioned tools are combined with the extended NeuroSim simulator to obtain the total power of the accelerators. Table~\ref{tab:ReDy_characteristics} shows the main characteristics of a single ReDy unit.

\begin{table}[t!]
\caption{Parameters of a ReDy Unit.}
\label{tab:ReDy_characteristics}
\centering
\resizebox{0.55\columnwidth}{!}{%
    \centering
    \begin{tabular}{|c|c|}
    \hline
    \cellcolor[gray]{0.9} Area & $768.8 um^2$ \\
    \cellcolor[gray]{0.9} Latency & $2.19 ns$ \\
    \cellcolor[gray]{0.9} Static Power & $9.3 uw$ \\
    \cellcolor[gray]{0.9} Dynamic Power & $62.9 uw$ \\
    \hline
    \end{tabular}%
}
\end{table}

Our objective is to demonstrate that our scheme provides significant energy savings for different CNN models. To this end, we evaluate our technique on three state-of-the-art networks for image classification including VGG-16~\cite{VGG}, ResNet-50~\cite{ResNet}, and DenseNet-161~\cite{DenseNet}. These networks are trained and evaluated with the ImageNet~\cite{ImageNet} dataset. Moreover, all the network models and our scheme have been implemented at software level with the PyTorch~\cite{Pytorch} and TensorFlow~\cite{Tensorflow} frameworks. For all the CNNs, we employ the whole test or validation dataset to obtain the top-1 accuracy metric, including several images to assess the efficiency of our dynamic quantization scheme in terms of accuracy loss. On the other hand, to reduce simulation time, we only use an statistically representative subset of the whole dataset to generate the traces of NeuroSim that are used to evaluate the energy consumption of the ReDy accelerator.

\section{Experimental Results}\label{s:Experimental_Results}
This section evaluates the accuracy, performance and energy efficiency of our dynamic quantization method. First, we introduce an analysis of the average numerical precision of activations and its impact in the accuracy of neural networks after applying the ReDy scheme. Next, we present the speedup and energy savings achieved by ReDy compared to a baseline ISAAC-like accelerator. In order to do a fair comparison, we evaluate the baseline and ReDy with the same floorplan configuration. Finally, we discuss the ReDy accelerator overheads.

\subsection{ReDy Evaluation}\label{subs:ReDy_eval}
Table~\ref{tab:Precision_Breakdown} shows the breakdown of the percentage of activations that are quantized to each numerical precision after applying ReDy. Our scheme provides very low bit-widths for the three CNNs that range from 5.7-bits (\textit{DenseNet-161}) to 6.8-bits (\textit{ResNet-50}), achieving an average quantization bitwidth of 6.1-bits. This reduction in bit-width results in a 33.8\% reduction in the activity of ReRAM crossbars and ADCs on average. The reduction in numerical precision and activity is due to our algorithm to determine the optimal precision of each group of activations on-the-fly, which takes into account the group's distribution to minimize the impact in accuracy loss. Furthermore, the overhead of performing the dynamic quantization is fairly small, since it is done concurrently while doing the analog computations. Table~\ref{tab:Accuracy_comparison} shows the impact on CNN accuracy where the average loss is lower than 0.85\%.

\begin{table}[t!]
\caption{Numerical precision breakdown of activations, average bitwidth, and ReRAM crossbars/ADCs activity reduction for VGG-16, ResNet-50 and DenseNet-161 CNNs.}
\label{tab:Precision_Breakdown}
\centering
\resizebox{1.0\columnwidth}{!}{%
    \centering
    \begin{tabular}{|c|ccc|}
    \hline
    Numerical Precision & VGG-16 & ResNet-50 & DenseNet-161 \\
    \hline
    8-bits &  1.6\% & 1.4\% & 0.5\% \\
    7-bits & 0\% & 78.8\% & 0\%    \\
    6-bits & 83.9\% & 19.6\% & 80.3\%  \\
    5-bits & 13.5\% & 0.1\% & 17.6\% \\
    Equal or Less than 4-bits & 1\% & 0.1\% & 1.6\% \\
    \hline\hline
    Average Bitwidth & 5.8-bits & 6.8-bits & 5.7-bits \\
    \hline\hline
    Activity Reduction & 42.2\% & 21.8\% & 37.3\% \\
    \hline
    \end{tabular}%
}
\end{table}

\begin{table}[t!]
\caption{Accuracy comparison for each CNN.}
\label{tab:Accuracy_comparison}
\centering
\resizebox{0.85\columnwidth}{!}{%
    \centering
    \begin{tabular}{|c|ccc|}
    \hline
    Top-1 Accuracy & VGG-16 & ResNet-50 & DenseNet-161 \\
    \hline
    Baseline & 71.57\% & 75.59\% & 77.07\% \\
    ReDy & 70.67\% & 74.97\% & 76.42\% \\
    \hline
    \end{tabular}%
}
\end{table}

\begin{figure}[t!]
    \centering
    \includegraphics[width=1.0\columnwidth]{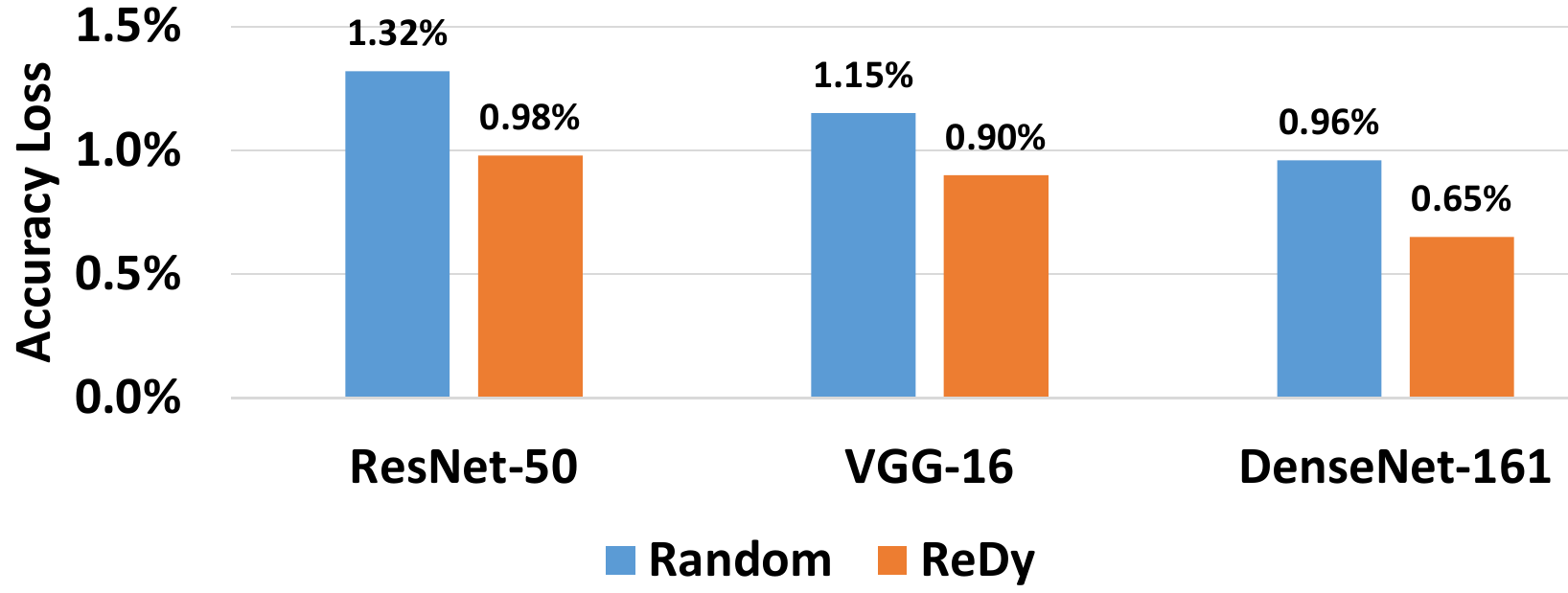}
    \caption{Accuracy loss of ReDy compared to a Random scheme with the same numerical precision breakdown of activations. Note that both quantization schemes have the same average bitwidth for a given CNN.}
    \label{fig:Accuracy_loss_over_random}
\end{figure}

To demonstrate the efficiency of ReDy in determining the best precision for each group of activations, Figure~\ref{fig:Accuracy_loss_over_random} shows the accuracy loss of ReDy compared to a random scheme with the same numerical precision breakdown of activations. Thus, in these experiments, the random quantization scheme employs the same overall average bitwidth as ReDy in each layer of a given CNN, but the precision of each group is randomly selected from a list of all the possible candidates obtained with the heuristic of ReDy. On average, the random scheme has an accuracy loss of 1.15\% whereas that of ReDy is 0.85\%, that is, ReDy exhibits a lower impact in accuracy loss when the average bitwidth is the same for both quantization schemes. Therefore, these results prove the sensibility to the distribution of each group of activations as an important criterion to assess the optimal quantization precision.

Similar to previous works~\cite{Neurosim_mapping, ISAAC}, all the layers of a CNN are executed concurrently in a deep pipeline to maximize the ReRAM crossbars utilization. Therefore, the ReDy accelerator throughput is dominated by the maximum latency among all the layers executed in the pipeline. For a given CONV layer, ReDy can reduce the latency required for the analog dot-product computations but, as described in Section~\ref{s:dq_method}, not all layers are quantized with ReDy, such as the first CONV layer of most CNNs, due to the small number of input channels. In some cases, this becomes the major limitation to improve the performance of the ReDy accelerator. Our results show that the throughput of DenseNet-161 is restricted by a layer where ReDy is applied and, hence, achieves an speedup of $1.33x$. On the other hand, VGG-16 and ResNet-50 are limited by the latency of the first CONV layer. Accordingly, ReDy cannot improve the performance of these networks. Note that there is no latency overhead in any case. In addition, ReDy can be customized according to the user preferences in terms of accuracy, performance and energy consumption. In future works, we plan to optimize the pipelined execution of the accelerator to exploit the full benefits of ReDy in terms of performance.

\begin{figure}[t!]
    \centering
    \includegraphics[width=1.0\columnwidth]{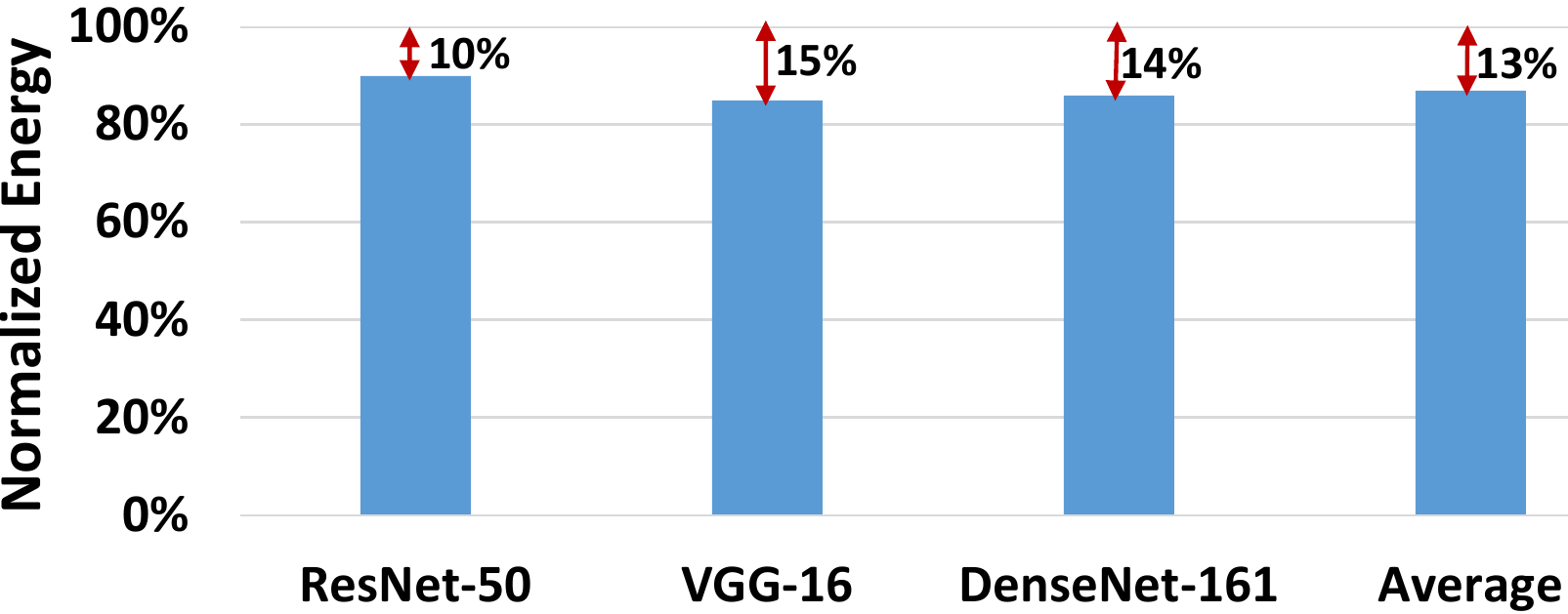}
    \caption{Normalized energy for each CNN. Baseline configuration is the ISAAC-like accelerator without the ReDy scheme.}
    \label{fig:Energy_Improvement}
\end{figure}

Figure~\ref{fig:Energy_Improvement} reports normalized energy. On average, ReDy reduces the energy consumption of the accelerator by 13\%. The energy savings are well correlated with the low quantization bitwidth of activations reported in Table~\ref{tab:Precision_Breakdown}. These energy savings are due to two main reasons. First, dynamic energy is reduced due to the savings in the activity of ReRAM crossbars and ADCs. Second, due to the performance limitations described above, static energy cannot be reduced as much as the dynamic energy, reducing the overall energy savings. Again, \textit{VGG-16} and \textit{DenseNet-161} obtain the largest benefits, achieving a reduction of 15\% and 14\% in energy respectively.

\subsection{Why use Dynamic Quantization?}\label{subs:Dynamic Mechanism}
\begin{figure}[t!]
    \centering
    \includegraphics[width=1.0\columnwidth]{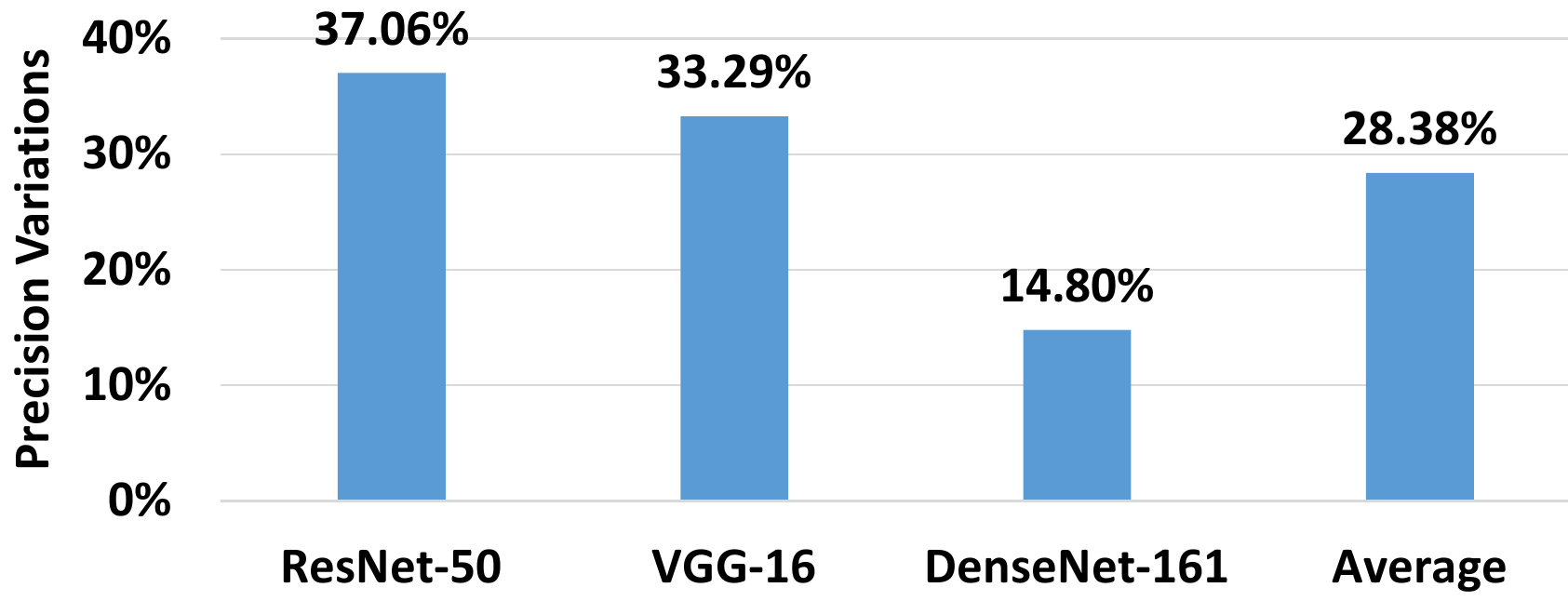}
    \caption{Percentage of non-zero variations in the numerical precision of the groups of activations between consecutive executions of the CNNs.}
    \label{fig:Dynamic_Mechanism}
\end{figure}

To verify the need for a dynamic quantization scheme such as ReDy, we compare the numerical precision of each group of activations in consecutive executions of the networks, that is, for a given group we compute the difference between the precision assigned by ReDy in the current execution and the previous one. Then, we accumulate the non-zero variations of all groups for multiple executions. Figure~\ref{fig:Dynamic_Mechanism} shows the percentage of accumulated non-zero variations when executing the entire validation dataset of each CNN. On average, we observe that there is a 28.38\% of variation in the numerical precision of the groups of activations. Therefore, ReDy can determine a distinct precision per group based on the proposed heuristic, and the selected precision for a given group does not converge to a specific value in consecutive executions of the neural network. As a result, ReDy benefits from a trade-off between energy savings and accuracy loss through a dynamic analysis of the activations.

\subsection{Overheads}\label{subs:overheads}
As shown in Table~\ref{tab:ReDy_characteristics}, ReDy requires extra hardware at the chip-level of the accelerator to determine the optimal numerical precision of each group of activations. The ReDy units are small compared to the ReRAM crossbars, ADCs, and SRAM buffers, so they represent a small portion of the area and energy of the accelerator. ReDy has a 0.20\% increase in area compared to the baseline ISAAC-like accelerator. On average, the energy consumed by the additional hardware of ReDy represents less than 0.8\% of the total energy. We believe ReDy's area and energy overheads are acceptable considering the energy improvements as reported in Section~\ref{subs:ReDy_eval}.

\section{Related Work}\label{s:Related_works}
This section outlines state-of-the-art ReRAM-based accelerators for DNNs and techniques to optimize these accelerators.

\textbf{PIM ReRAM-based Accelerators.} In recent years, PIM architectures~\cite{marc} have regained popularity and attracted the attention of researchers due to the high memory requirements of modern machine learning applications and the limitations of traditional systems. Implementing DNN accelerators using memristors is quite effective and, hence, most recent works are leveraging the use of ReRAM crossbars to accelerate cognitive computing algorithms~\cite{ISAAC, PRIME, PUMA, CASCADE, MNEMOSENE, PipeLayer, RAPIDNN, RACER, LADDER, MemUniSon, FPSA}. PRIME~\cite{PRIME} and ISAAC~\cite{ISAAC} are two of the most popular PIM accelerators for DNN/CNN inference with in-situ analog arithmetic in ReRAM crossbars. In PRIME, ReRAM banks are divided into memory (Mem), buffer and full-function (FF) subarrays. Mem subarrays only store data whereas FF subarrays can either store data or perform computations. On the other hand, ISAAC proposes a deep pipelined architecture with new strategies and data encoding techniques that are more amenable to analog computation. To addresses the problems of high-cost A/D conversions, CASCADE~\cite{CASCADE} proposes to extend the processing in analog domain and in-memory. In contrast, RAPIDNN~\cite{RAPIDNN} supports all DNN functionalities in a digital-based memory design. In a similar line of research, RACER~\cite{RACER} proposes a cost-effective PIM architecture that delivers high performance and energy savings making use of a bit-pipelining execution model with small arrays of ReRAM. Our work is different from these previous accelerators as we propose a dynamic quantization scheme of the activations to reduce the activity in the ReRAM crossbars and ADCs, which consume most of the power of ReRAM-based accelerators.

\textbf{ReRAM-centric Quantization and Pruning.} Similar to previous DNN accelerators, quantization and pruning are popular optimizations to improve the performance and energy efficiency of ReRAM-based architectures~\cite{FORMS, Flex-PIM, Q-PIM, Trainig_quantize, SubMac, sparse_ReRAM, Tiny_but_Accurate, PQ-PIM}. FORMS~\cite{FORMS} proposes a quantization and mapping scheme with polarized weights that is performed during the training of the DNN model to achieve high accuracy. Their key idea is to enforce that all weights in the same column of a crossbar have the same sign to reduce the overheads of computing with signed weights. The work in \cite{Trainig_quantize} also considers training and quantization at the same time to generate a discrete model that alleviates the accuracy loss that occurs when mapping the quantized kernels and activations to the ReRAM crossbars. On the other hand, the Sparse ReRAM Engine~\cite{sparse_ReRAM} exploits both the weight and activation bit sparsity resulting from pruning to reduce the activity of ReRAM crossbars and ADCs. However, the pruned models require re-training to recover the full accuracy. The authors of \cite{Tiny_but_Accurate} propose a ReRAM-based DNN framework that combines structured weight pruning and static quantization to improve performance. Finally, SubMac~\cite{SubMac} proposes a quantization scheme to skip subsets of computations based on the magnitude of inputs and weights. To the best of our knowledge, ReDy is the first proposal that exploits the bit-serial execution within crossbars to perform on-the-fly quantization without re-training based on the distribution of activation groups in CNNs.

\section{Conclusion}\label{s:conclusion}
In this paper, we show that modern CNNs exhibit a high precentage of activation groups with a close to uniform distribution in CONV layers, resulting in a minor impact in accuracy when quantized to low bitwidths. Then, we propose ReDy, a new ReRAM-based PIM accelerator that exploits the bit-serial processing of groups of activations to reduce the ReRAM crossbars activity and the number of A/D conversions. ReDy implements a dynamic quantization scheme where each group of activations is quantized on-the-fly to a different numerical precision based on the distribution of activations in the group and its similarity to a uniform distribution. We show that ReDy requires minor hardware changes over a state-of-the-art accelerator, mainly additional counters and comparators to compute the histogram of activations and some functional units to estimate the error. Our experimental results show that, on average, ReDy provides a 33.8\% reduction in the activity of the ReRAM crossbars and ADCs which brings 13\% energy savings with negligible accuracy loss, while it only requires a minor increase in the area of the accelerator (less than 0.20\%).

\section{Acknowledgments}\label{s:Acknowledgments}
This work has been supported by the CoCoUnit ERC Advanced Grant of the EU’s Horizon 2020 program (grant No 833057), the Spanish State Research Agency (MCIN/AEI) under grant PID2020-113172RB-I00, and the ICREA Academia program.


\bibliographystyle{IEEEtran}
\bibliography{refs}

\begin{thebibliography}{10}
\providecommand{\url}[1]{#1}
\csname url@samestyle\endcsname
\providecommand{\newblock}{\relax}
\providecommand{\bibinfo}[2]{#2}
\providecommand{\BIBentrySTDinterwordspacing}{\spaceskip=0pt\relax}
\providecommand{\BIBentryALTinterwordstretchfactor}{4}
\providecommand{\BIBentryALTinterwordspacing}{\spaceskip=\fontdimen2\font plus
\BIBentryALTinterwordstretchfactor\fontdimen3\font minus
  \fontdimen4\font\relax}
\providecommand{\BIBforeignlanguage}[2]{{%
\expandafter\ifx\csname l@#1\endcsname\relax
\typeout{** WARNING: IEEEtran.bst: No hyphenation pattern has been}%
\typeout{** loaded for the language `#1'. Using the pattern for}%
\typeout{** the default language instead.}%
\else
\language=\csname l@#1\endcsname
\fi
#2}}
\providecommand{\BIBdecl}{\relax}
\BIBdecl

\bibitem{data_transfer_mobile}
D.~Pandiyan and C.-J. Wu, ``Quantifying the energy cost of data movement for
  emerging smart phone workloads on mobile platforms,'' in \emph{IEEE
  International Symposium on Workload Characterization (IISWC)}, 2014, pp.
  171--180.

\bibitem{data_transfer_totall}
G.~Kestor, R.~Gioiosa, D.~J. Kerbyson, and A.~Hoisie, ``Quantifying the energy
  cost of data movement in scientific applications,'' in \emph{IEEE
  International Symposium on Workload Characterization (IISWC)}, 2013, pp.
  56--65.

\bibitem{TPU}
N.~P. Jouppi, C.~Young, N.~Patil, D.~A. Patterson, G.~Agrawal, R.~Bajwa,
  S.~Bates, and et~al., ``In-datacenter performance analysis of a tensor
  processing unit,'' in \emph{Proceedings of the 44th Annual International
  Symposium on Computer Architecture}, 2017, p. 1–12.

\bibitem{MNEMOSENE}
M.~Zahedi, M.~A. Lebdeh, C.~Bengel, D.~Wouters, S.~Menzel, M.~Le~Gallo,
  A.~Sebastian, S.~Wong, and S.~Hamdioui, ``Mnemosene: Tile architecture and
  simulator for memristor-based computation-in-memory,'' \emph{ACM Journal on
  Emerging Technologies in Computing Systems}, vol.~18, no.~3, 2022.

\bibitem{PUMA}
A.~Ankit, I.~E. Hajj, S.~R. Chalamalasetti, G.~Ndu, M.~Foltin, R.~S. Williams,
  P.~Faraboschi, W.-m.~W. Hwu, J.~P. Strachan, K.~Roy, and D.~S. Milojicic,
  ``Puma: A programmable ultra-efficient memristor-based accelerator for
  machine learning inference,'' in \emph{Proceedings of the Twenty-Fourth
  International Conference on Architectural Support for Programming Languages
  and Operating Systems (ASPLOS)}, 2019, p. 715–731.

\bibitem{FORMS}
G.~Yua, P.~Behnam, Z.~Li, A.~Shafiee, S.~Lin, X.~Ma, H.~Liu, X.~Qian,
  M.~Bojnordi, Y.~Wang, and C.~Ding, ``Forms: Fine-grained polarized
  reram-based in-situ computation for mixed-signal dnn accelerator,'' in
  \emph{ACM/IEEE 48th Annual International Symposium on Computer Architecture
  (ISCA)}, 2021, pp. 265--278.

\bibitem{ISAAC}
A.~Shafiee, A.~Nag, N.~Muralimanohar, R.~Balasubramonian, J.~P. Strachan,
  M.~Hu, R.~S. Williams, and V.~Srikumar, ``Isaac: A convolutional neural
  network accelerator with in-situ analog arithmetic in crossbars,'' in
  \emph{ACM/IEEE 43rd Annual International Symposium on Computer Architecture
  (ISCA)}, 2016, pp. 14--26.

\bibitem{Direct-Current-Free}
J.-M. Hung, T.-H. Wen, Y.-H. Huang, S.-P. Huang, F.-C. Chang, C.-I. Su, W.-S.
  Khwa, C.-C. Lo, R.-S. Liu, C.-C. Hsieh, K.-T. Tang, Y.-D. Chih, T.-Y.~J.
  Chang, and M.-F. Chang, ``8-b precision 8-mb reram compute-in-memory macro
  using direct-current-free time-domain readout scheme for ai edge devices,''
  \emph{IEEE Journal of Solid-State Circuits}, vol.~58, 2023.

\bibitem{ResiRCA}
K.~Qiu, N.~Jao, M.~Zhao, C.~S. Mishra, G.~Gudukbay, S.~Jose, J.~Sampson, M.~T.
  Kandemir, and V.~Narayanan, ``Resirca: A resilient energy harvesting reram
  crossbar-based accelerator for intelligent embedded processors,'' in
  \emph{2020 IEEE International Symposium on High Performance Computer
  Architecture (HPCA)}, 2020.

\bibitem{XNOR-RRAM}
X.~Sun, S.~Yin, X.~Peng, R.~Liu, J.-s. Seo, and S.~Yu, ``Xnor-rram: A scalable
  and parallel resistive synaptic architecture for binary neural networks,'' in
  \emph{Design, Automation \& Test in Europe Conference \& Exhibition (DATE)},
  2018.

\bibitem{CASCADE}
T.~Chou, W.~Tang, J.~Botimer, and Z.~Zhang, ``Cascade: Connecting rrams to
  extend analog dataflow in an end-to-end in-memory processing paradigm,'' in
  \emph{Proceedings of the 52nd Annual IEEE/ACM International Symposium on
  Microarchitecture}, 2019, p. 114–125.

\bibitem{PRIME}
P.~Chi, S.~Li, C.~Xu, T.~Zhang, J.~Zhao, Y.~Liu, Y.~Wang, and Y.~Xie, ``Prime:
  A novel processing-in-memory architecture for neural network computation in
  reram-based main memory,'' in \emph{ACM/IEEE 43rd Annual International
  Symposium on Computer Architecture (ISCA)}, 2016, pp. 27--39.

\bibitem{PCM}
B.~C. Lee, E.~Ipek, O.~Mutlu, and D.~Burger, ``Phase change memory architecture
  and the quest for scalability,'' \emph{Communications of the ACM}, vol.~53,
  no.~7, p. 99–106, 2010.

\bibitem{STT}
M.~Hosomi, H.~Yamagishi, T.~Yamamoto, K.~Bessho, Y.~Higo, K.~Yamane, H.~Yamada,
  M.~Shoji, H.~Hachino, C.~Fukumoto, H.~Nagao, and H.~Kano, ``A novel
  nonvolatile memory with spin torque transfer magnetization switching:
  spin-ram,'' in \emph{IEEE International Electron Devices Meeting (IEDM)},
  2005, pp. 459--462.

\bibitem{MPIM}
M.~Imani, Y.~Kim, and T.~Rosing, ``Mpim: Multi-purpose in-memory processing
  using configurable resistive memory,'' in \emph{22nd Asia and South Pacific
  Design Automation Conference (ASP-DAC)}, 2017, pp. 757--763.

\bibitem{Energy_Efficient}
J.~Liu, H.~Zhao, M.~A. Ogleari, D.~Li, and J.~Zhao, ``Processing-in-memory for
  energy-efficient neural network training: A heterogeneous approach,'' in
  \emph{51st Annual IEEE/ACM International Symposium on Microarchitecture
  (MICRO)}, 2018, pp. 655--668.

\bibitem{Boolean}
A.~Siemon, D.~Wouters, S.~Hamdioui, and S.~Menzel, ``Memristive device modeling
  and circuit design exploration for computation-in-memory,'' in \emph{IEEE
  International Symposium on Circuits and Systems (ISCAS)}, 2019, pp. 1--5.

\bibitem{ResNet}
K.~He, X.~Zhang, S.~Ren, and J.~Sun, ``Deep residual learning for image
  recognition,'' \emph{arXiv}, 2015.

\bibitem{Neurosim_trend}
S.~Yu, H.~Jiang, S.~Huang, X.~Peng, and A.~Lu, ``Compute-in-memory chips for
  deep learning: Recent trends and prospects,'' \emph{IEEE Circuits and Systems
  Magazine}, vol.~21, no.~3, pp. 31--56, 2021.

\bibitem{survey_Qauntization}
A.~Gholami, S.~Kim, Z.~Dong, Z.~Yao, M.~W. Mahoney, and K.~Keutzer, ``A survey
  of quantization methods for efficient neural network inference,''
  \emph{arXiv}, 2021.

\bibitem{marc}
M.~Hassanpour, M.~Riera, and A.~Gonz{\'a}lez, ``A survey of near-data
  processing architectures for neural networks,'' \emph{Machine Learning and
  Knowledge Extraction}, 2022.

\bibitem{Neurosim_mapping}
X.~Peng, R.~Liu, and S.~Yu, ``Optimizing weight mapping and data flow for
  convolutional neural networks on rram based processing-in-memory
  architecture,'' in \emph{IEEE International Symposium on Circuits and Systems
  (ISCAS)}, 2019, pp. 1--5.

\bibitem{Training_map}
T.~Gokmen, O.~M. Onen, and W.~Haensch, ``Training deep convolutional neural
  networks with resistive cross-point devices,'' \emph{Frontiers in
  Neuroscience}, vol.~11, 2017.

\bibitem{dynamic_reram}
X.~Chen, J.~Jiang, J.~Zhu, and C.-Y. Tsui, ``A high-throughput and
  energy-efficient rram-based convolutional neural network using data encoding
  and dynamic quantization,'' in \emph{23rd Asia and South Pacific Design
  Automation Conference (ASP-DAC)}, 2018, pp. 123--128.

\bibitem{Neurosim_github}
G.~Tech, ``Neurosim,'' \url{https://github.com/neurosim/}.

\bibitem{Design_compiler}
Synopsys, ``Design compiler,'' \url{https://www.synopsys.com/}.

\bibitem{ModelSim}
M.~Graphics, ``Modelsim,'' \url{https://www.intel.com/}.

\bibitem{VGG}
K.~Simonyan and A.~Zisserman, ``Very deep convolutional networks for
  large-scale image recognition,'' \emph{arXiv}, 2014.

\bibitem{DenseNet}
G.~Huang, Z.~Liu, L.~van~der Maaten, and K.~Q. Weinberger, ``Densely connected
  convolutional networks,'' \emph{arXiv}, 2016.

\bibitem{ImageNet}
S.~V. Lab, ``Imagenet,'' \url{https://www.image-net.org/}.

\bibitem{Pytorch}
Meta, ``Pytorch,'' \url{https://pytorch.org/}.

\bibitem{Tensorflow}
Google, ``Tensorflow,'' \url{https://www.tensorflow.org/}.

\bibitem{PipeLayer}
L.~Song, X.~Qian, H.~Li, and Y.~Chen, ``Pipelayer: A pipelined reram-based
  accelerator for deep learning,'' in \emph{IEEE International Symposium on
  High Performance Computer Architecture (HPCA)}, 2017, pp. 541--552.

\bibitem{RAPIDNN}
M.~Imani, M.~Samragh~Razlighi, Y.~Kim, S.~Gupta, F.~Koushanfar, and T.~Rosing,
  ``Deep learning acceleration with neuron-to-memory transformation,'' in
  \emph{IEEE International Symposium on High Performance Computer Architecture
  (HPCA)}, 2020, pp. 1--14.

\bibitem{RACER}
M.~S.~Q. Truong, E.~Chen, D.~Su, L.~Shen, A.~Glass, L.~R. Carley, J.~A. Bain,
  and S.~Ghose, ``Racer: Bit-pipelined processing using resistive memory,'' in
  \emph{54th Annual IEEE/ACM International Symposium on Microarchitecture
  (MICRO)}, 2021, p. 100–116.

\bibitem{LADDER}
M.~H.~I. Chowdhuryy, M.~R.~H. Rashed, A.~Awad, R.~Ewetz, and F.~Yao, ``Ladder:
  Architecting content and location-aware writes for crossbar resistive
  memories,'' in \emph{54th Annual IEEE/ACM International Symposium on
  Microarchitecture (MICRO)}, 2021, p. 117–130.

\bibitem{MemUniSon}
J.~Wang, J.~Liu, D.~Wang, S.~Zhang, and X.~Fan, ``Memunison: A
  racetrack-reram-combined pipeline architecture for energy-efficient in-memory
  cnns,'' \emph{IEEE Transactions on Computers}, vol.~71, 2022.

\bibitem{FPSA}
Y.~Ji, Z.~Liu, and Y.~Zhang, ``A reduced architecture for reram-based neural
  network accelerator and its software stack,'' \emph{IEEE Transactions on
  Computers}, vol.~70, 2021.

\bibitem{Flex-PIM}
Y.~Long, E.~Lee, D.~Kim, and S.~Mukhopadhyay, ``Flex-pim: A ferroelectric fet
  based vector matrix multiplication engine with dynamical bitwidth and
  floating point precision,'' in \emph{International Joint Conference on Neural
  Networks (IJCNN)}, 2020, pp. 1--8.

\bibitem{Q-PIM}
------, ``Q-pim: A genetic algorithm based flexible dnn quantization method and
  application to processing-in-memory platform,'' in \emph{57th ACM/IEEE Design
  Automation Conference (DAC)}, 2020, pp. 1--6.

\bibitem{Trainig_quantize}
Q.~Yang, H.~Li, and Q.~Wu, ``A quantized training method to enhance accuracy of
  reram-based neuromorphic systems,'' in \emph{IEEE International Symposium on
  Circuits and Systems (ISCAS)}, 2018, pp. 1--5.

\bibitem{SubMac}
X.~Chen, J.~Jiang, Z.~Jingyang, and C.-y. Tsui, ``Submac: Exploiting the
  subword-based computation in rram-based cnn accelerator for energy saving and
  speedup,'' \emph{Integration}, vol.~69, pp. 356--368, 09 2019.

\bibitem{sparse_ReRAM}
T.-H. Yang, H.-Y. Cheng, C.-L. Yang, I.-C. Tseng, H.-W. Hu, H.-S. Chang, and
  H.-P. Li, ``Sparse reram engine: Joint exploration of activation and weight
  sparsity in compressed neural networks,'' in \emph{ACM/IEEE 46th Annual
  International Symposium on Computer Architecture (ISCA)}, 2019, pp. 236--249.

\bibitem{Tiny_but_Accurate}
X.~Ma, G.~Yuan, S.~Lin, C.~Ding, F.~Yu, T.~Liu, W.~Wen, X.~Chen, and Y.~Wang,
  ``Tiny but accurate: A pruned, quantized and optimized memristor crossbar
  framework for ultra efficient dnn implementation,'' in \emph{2020 25th Asia
  and South Pacific Design Automation Conference (ASP-DAC)}, 2020, pp.
  301--306.

\bibitem{PQ-PIM}
Z.~Yuhao, W.~Xinyu, J.~Xikun, Y.~Yuhan, S.~Zhaoyan, and J.~Zhiping, ``Q-pim: A
  pruning-quantization joint optimization framework for reram-based
  processing-in-memory {DNN} accelerator,'' \emph{J. Syst. Archit.}, vol. 127,
  2022.

\end{thebibliography}


\begin{IEEEbiography}[{\includegraphics[width=1in,height=1.25in,clip,keepaspectratio]{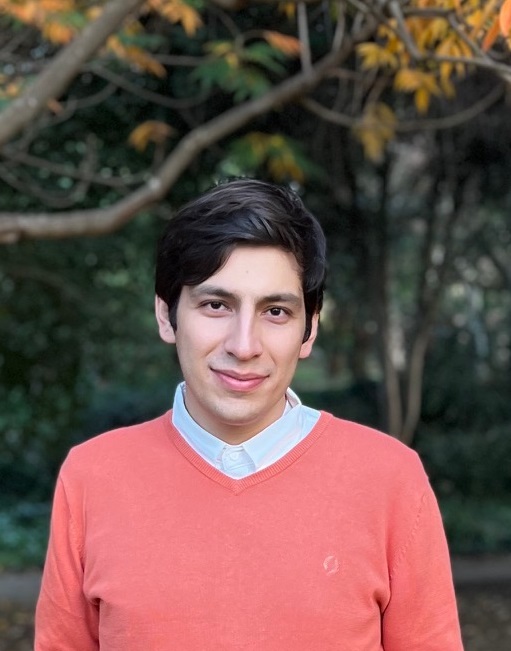}}]{Mohammad Sabri}
received a B.S. degree in electrical engineering from Shahid Rajaee University, Tehran, Iran, in 2015, and an M.S. degree in digital electronic systems at Tehran University, Tehran, Iran in 2019. He is currently PhD researcher at Universitat Politècnica de Catalunya (UPC - BarcelonaTech) at ARCO group. His research interests include Near-memory computing, and cognitive computing systems.
\end{IEEEbiography}

\begin{IEEEbiography}[{\includegraphics[width=1in,height=1.25in,clip,keepaspectratio]{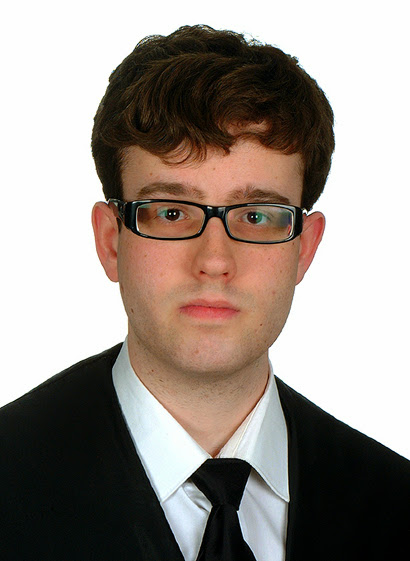}}]{Marc Riera}
received his B.S degree in Computer Engineering in 2013, his MS degree in MIRI: High Performance Computing in 2015, and his PhD in Computer Architecture in 2020, all from Universitat Politècnica de Catalunya (UPC - BarcelonaTech). He joined the ARCO research group in 2014, and is currently a postdoctoral researcher at ARCO. His research interests focus on the area of Accelerator Architectures, Machine Learning, and Near-Data Processing (NDP).
\end{IEEEbiography}

\begin{IEEEbiography}[{\includegraphics[width=1in,height=1.25in,clip,keepaspectratio]{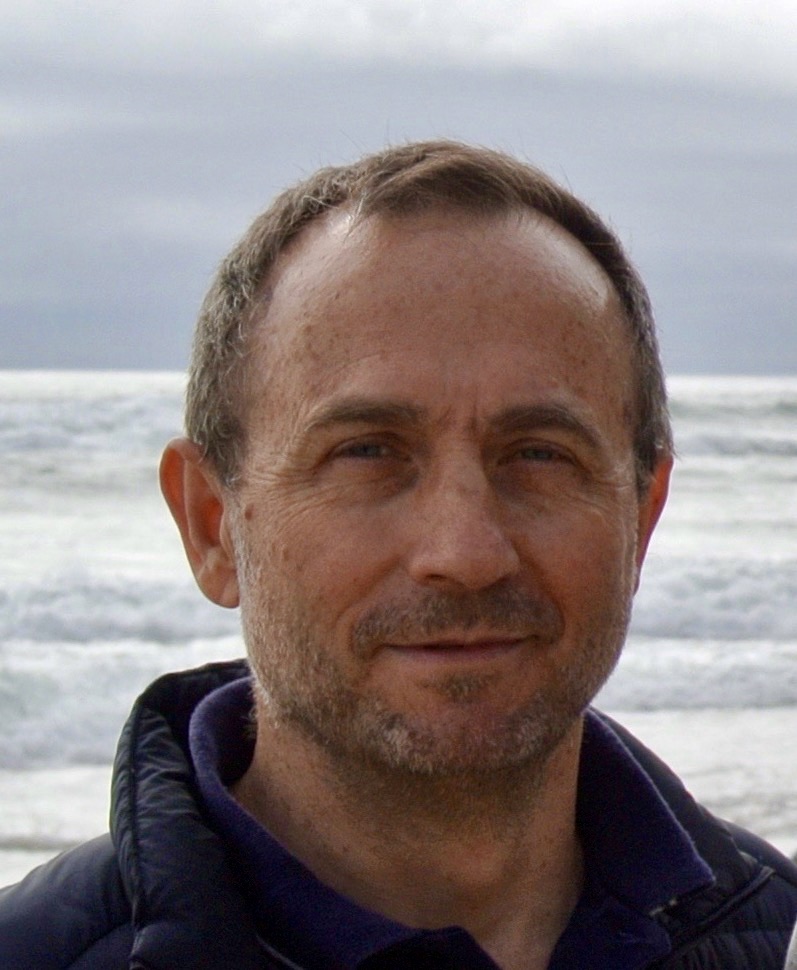}}]{Antonio~Gonz\'alez}
(PhD 1989) is a Full Professor at the Computer Architecture Department of the Universitat Politècnica de Catalunya, Barcelona (Spain), and the director of the Architecture and Compilers research group. He was the founding director of the Intel Barcelona Research Center from 2002 to 2014. His research has focused on computer architecture and compilers, with a special emphasis on cognitive computing systems and graphics processors in recent years. He has published over 400 papers, and has served as associate editor of five IEEE and ACM journals, program chair for ISCA, MICRO, HPCA, ICS and ISPASS, and general chair for MICRO and HPCA. He is a Fellow of IEEE and ACM.
\end{IEEEbiography}

\vfill

\end{document}